\documentclass[acmsmall,manuscript]{acmart}
\usepackage{graphicx}
\usepackage{multirow}
\usepackage{array}
\usepackage{longtable}
\usepackage[figuresright]{rotating}
\usepackage{enumitem}
\usepackage{caption,subcaption}
\usepackage{color,soul}
\usepackage{float}
\usepackage{xcolor} 
\usepackage{tabularx}
\usepackage{colortbl}
\newcommand\crule[3][black]{\textcolor{#1}{\rule{#2}{#3}}}

\AtBeginDocument{%
  \providecommand\BibTeX{{%
    \normalfont B\kern-0.5em{\scshape i\kern-0.25em b}\kern-0.8em\TeX}}}

\setcopyright{acmcopyright}
\copyrightyear{2024}
\acmYear{2024}
\acmDOI{10.1145/1122445.1122456}

\acmJournal{JACM}
\acmVolume{37}
\acmNumber{4}
\acmArticle{111}
\acmMonth{6}

\begin{document}

\title{Visualization for Recommendation Explainability: A Survey and New Perspectives}


\author{Mohamed Amine Chatti}
\affiliation{%
  \institution{University of Duisburg-Essen}
  \city{Duisburg}
  \country{Germany}}
\email{mohamed.chatti@uni-due.de}

\author{Mouadh Guesmi}
\affiliation{%
  \institution{University of Duisburg-Essen}
\city{Duisburg}
\country{Germany}}
\email{mouadh.guesmi@stud.uni-due.de}

\author{Arham Muslim}
\affiliation{%
  \institution{National University of Sciences and Technology}
\city{Islamabad}
\country{Pakistan}}
\email{arham.muslim@seecs.edu.pk}

\renewcommand{\shortauthors}{Chatti, et al.}

\begin{abstract}
Providing system-generated explanations for recommendations represents an important step towards transparent and trustworthy recommender systems. 
Explainable recommender systems provide a human-understandable rationale for their outputs.
Over the past two decades, explainable recommendation has attracted much attention in the recommender systems research community.
This paper aims to provide a comprehensive review of research efforts on visual explanation in recommender systems. More concretely, we systematically review the literature on explanations in recommender systems based on four dimensions, namely explanation aim, explanation scope, explanation method, and explanation format. Recognizing the importance of visualization, we approach the recommender system literature from the angle of explanatory visualizations, that is using visualizations as a display style of explanation. As a result, we derive a set of guidelines that might be constructive for designing explanatory visualizations in recommender systems and identify perspectives for future work in this field. The aim of this review is to help recommendation researchers and practitioners better understand the potential of visually explainable recommendation research and to support them in the systematic design of visual explanations in current and future recommender systems.
\end{abstract}

\ccsdesc[500]{Information systems~Recommender systems}
\ccsdesc[500]{Human-centered computing~Visualization techniques}
\ccsdesc[500]{Human-centered computing~Information visualization}
\ccsdesc[500]{Human-centered computing~Interactive systems and tools}

\keywords{recommender system, explainable recommendation, visualization}

\maketitle

\section{Introduction}
Current recommender systems often appear as \textit{black box} to their users by hiding important details about why items are recommended and how they relate to the users’ preferences. The lack of transparency has sparked increased interest in incorporating explanation in recommender systems, with the goal of making these systems more transparent and providing users with information that can aid in the development of an accurate mental model of the system's behavior \cite{zhang2020explainable, Nunes2017, Tintarev2015}. Explainable recommendation refers to personalized recommendation algorithms that not only provide users with recommendations, but also provide an explanation for why a specific item has been recommended \cite{Herlocker2000, Bilgic2005, Tintarev2007, Friedrich2011}.
Visualization techniques can provide a visual entry to effectively explain the working of a recommender system. 
The human visual system is the highest-bandwidth channel into the human brain \cite{Ware2012}. Humans, in general, can process visual information faster and much easier as compared to textual information \cite{Munzner2014}. The purpose of visualization is insight \cite{north2006toward}. By visualizing data, some insights emerge which might not be noticeable in the raw form of the data \cite{hearst2009search}. Moreover, visualization leverages the functioning of the human visual system to provide a powerful means of making sense of data \cite{heer2010tour}, to communicate complex information \cite{ward2010interactive}, to help humans resolve logical problems, to think and reason \cite{card1999readings}, and to support decision making \cite{padilla2018decision}.
%
%
%
Thus, a natural way to obtain human-interpretable explanations is through visualizations \cite {spinner2019explainer}, making them a popular medium to provide insight into data or how a system works \cite{lim2009why}. Another benefit of visualizations is that they can provide a viable path to interactive explanations; an iterative process, where the system explains and users can give feedback and correct the system’s assumptions if necessary \cite{jannach2019explanations}. Interactive visualization can be an effective means of communication between the user and the recommender system and a powerful tool for understanding and exploring how the recommender system works.  

The role of visualization in explainable artificial intelligence (XAI) gained significant attention in recent years to explain opaque AI models \cite{spinner2019explainer, yang2020visual}. Acknowledging their value, visualizations are also widely used to deliver explanations in recommender systems \cite{Nunes2017, zhang2020explainable}. Early works on explainable recommendation have proposed various ways of explaining recommendations to the user based on visualizations (e.g., bar chart, pie chart, histogram, tag cloud) and have shown that visual explanations can help to enhance transparency and users' trust, and improve the overall acceptance of a recommender system (see, e.g., \cite{Herlocker2000, Bilgic2005, Vig2009, Gedikli2014}). Since then, the body of research on adopting visualization as a format to explain recommender systems has continued to grow.   
Despite the fruitful research findings, there is still a lack of a clear picture in the literature on explainable recommendation regarding adopting visualization. Moreover, it becomes important to develop a better understanding on the potential use of visualization for recommendation explainability. 
While there has been a considerable amount of research on explainable recommendation and excellent systematic literature reviews of explainable recommender systems based on different dimensions \cite{zhang2020explainable, Nunes2017}, and visualizations in interactive recommendation \cite{he2016interactive,jugovac2017interacting}, approaching the recommender system literature from the angle of explanatory visualization, that is using visualization as a display style of explanation, is not very common. Moreover, designing effective visual explanations to help end users understand how an recommender system operates can be quite challenging. 
\citet{Gedikli2014} proposed a set of possible guidelines for designing or selecting suitable explanations for recommender systems, e.g., "Increase transparency through explanations for higher user satisfaction". 
\citet{kulesza2015principles} outlined a set of principles for designing explanations for interactive machine learning, e.g., "Be Sound", "Be Complete" and "Don’t Overwhelm". While these studies explored different design factors for explanations in recommender and machine learning systems, investigating visual design properties of explanations is lacking in the literature on explainable recommendation. 

To close these research gaps, we review in this paper the work on explainable recommendation that has been published in the past two decades, specifically between 2000 and 2021, with a focus on explanatory visualizations. We conduct a comprehensive survey of visually explainable recommendation based on four dimensions, namely explanation aim, scope, method, and format. Further, we describe guidelines to ground and support the design of effective visual explanations in recommender systems. The main aim of this work is to help recommender system researchers and practitioners get the big picture of explainable recommendation augmented by visualizations with the hope that they will have rough guideline when it comes to choosing visualizations to solve explainable recommendation tasks at hand.


The remainder of this paper is organized as follows: In Section \ref{sec:relwork}, we first outline the background for this research and three branches of related work, namely visual explanation, interactive recommendation, and interactive explanation. Moreover, we summarize several of the most relevant peer-reviewed surveys related to the topic of explainable recommendation. We then introduce the four classification dimensions used in this survey in Section \ref{sec:dim_exp}. Munzner's what-why-how visualization framework is briefly explained in Section \ref{subsec:vis}. This framework provides the theoretical background to discuss how explanatory visualizations are provided in the reviewed literature. Section \ref{sec:method} describes the steps that were taken while planning the survey. Section \ref{sec:result} then discusses in detail the results of this survey. Section \ref{sec:discussion} provides a set of possible design guidelines and future lines of work in this field. Section \ref{sec:conclusion} finally summarizes the main findings of this work.
\section{Background and Related Work} \label{sec:relwork}
Our survey touches upon explainable AI (XAI), explainable recommendation, visual explanation, visual analytics, interactive recommendation, and interactive explanation. This section presents relevant work at the intersection of these domains.
\subsection{Explainable Recommendation} \label{sec:exp}
Recommender systems represent an important branch of AI research. While, for a long time, research in recommender systems focused primarily on algorithmic accuracy, there has been an increased interest in more user-centric evaluation metrics such as the system's transparency and trustworthiness or the degree of control users are able to exert over the recommendation process \cite{Andjelkovic2019,  Knijnenburg2012, Konstan2012, Pu2011, Pu2012}. 
A major approach to enhance transparency in recommender systems is to provide the rationale behind a recommendation in the form of explanations \cite{Tintarev2015}. Explanations are a necessary condition to help users build an accurate mental model of the recommender system. A popular interpretation of the term explanation in recommender systems is that explanations ``justify'' the recommendations \cite{Tintarev2012} to help humans understand why certain items are recommended by the system. 
Explanations in recommender systems are often characterized through the possible aims which one might want to achieve with them. These include efficiency, effectiveness, persuasiveness, transparency, satisfaction, scrutability, and trust \cite{Tintarev2007}. 
Explanations in recommender systems also facilitate system designers to diagnose, debug, and refine the recommendation algorithm \cite{zhang2020explainable}. Generally, explanations seek to show how a recommended item relates to users' preferences \cite{Vig2009}. At a high level, an explanation is an answer to a question \cite{miller2019explanation}, also called intelligibility type, such as \textit{What}, \textit{Why}, \textit{How}, \textit{What if}, and \textit{Why not} \cite{lim2009assessing, liao2020questioning}, that leads to understanding.
The primary aim of an explanation is to tell users about the information the recommender system had at its disposal and how it used that information to provide recommendations. 

The explainability of recommendations has attracted considerable attention over the past two decades and many explainable recommendation approaches have been proposed \cite{Tintarev2015, zhang2020explainable, Nunes2017}. Research on explainable recommendation has focused on different dimensions, including explanation aim, method, scope, and format. We discuss these dimensions in detail in Section \ref{sec:dim_exp}. 
In terms of explanation format, prior work differentiates between textual and visual presentation of explanations. 
To provide a deeper understanding of the value of visual explanation, we can draw upon psychological and communication theories such as the dual-process theory \cite{kahneman2011thinking} and the elaboration likelihood model (ELM) \cite{petty1986elaboration}. These theories offer insights into how individuals process information and make decisions. The dual-process theory \cite{ kahneman2002representativeness,kahneman2011thinking,evans2013dual} informs the design of effective information visualizations by acknowledging the two distinct thinking systems humans possess: System 1 (i.e., the instant, unconscious, automatic, emotional, intuitive thinking) and System 2 (i.e., the slower, conscious, rational, reasoning, deliberate, analytical thinking). Visualizations can cater to both System 1, the fast, automatic processing system, and System 2, the slower, effortful. Leveraging pre-attentive processing features like position, length, color, size, shape, angle, and area (elements readily grasped by System 1), visualizations can transmit information quickly and intuitively.  Furthermore, visualizations can be designed to support System 2 processing by allowing users to interact with the data, thus enabling the induction of insight, reasoning, and understanding, as well as facilitating hypothesis generation and deeper analysis, which are key goals of the visual analytics community. This dual-process approach can optimize information visualization for both rapid comprehension and complex reasoning \cite{patterson2014human,padilla2018decision}.

\citet{patterson2014human} presented a human cognition framework for information visualization. The authors view information visualization fundamentally as a human cognition-augmentation issue and proposed that well-designed visualizations induce insight, reasoning, and understanding by leveraging human visual perception capabilities (i.e., System 1) and influencing high-level cognitive processes such as retrieval from long-term memory (i.e., System 2). In relation to decision making, \citet{padilla2018decision} proposed a framework for decision making with visualizations, which expands on theories of visualization cognition and decision making. This framework is in line with previous theories from the human cognition literature in asserting that visualization decision making is an integration of both bottom-up (i.e., System 1) and top-down (i.e., System 2) processing. Consistent with a dual-process model, this framework suggests that the first type of visualization decision mechanism produces fast, easy, and computationally light decisions with visualizations. The second facilitates slower, contemplative, and effortful decisions with visualizations.

Another prominent theoretical framework of interest in the context of explainable recommendation is the elaboration likelihood model (ELM), which represents a “dual process” approach that is focused specifically on persuasion \cite{petty1986elaboration}. While the dual-process theory provides a comprehensive framework for understanding cognitive processing through its distinction between automatic and controlled processes, the ELM further delves into the mechanisms underlying persuasion by elucidating the processes involved in attitude change. Building upon the foundation of dual-process theory, ELM offers insights into how the message recipient engages in “elaboration” (i.e., degree of thought) about information relevant to the persuasive topic. The degree of elaboration forms a continuum from extremely high elaboration to little or no elaboration. The ELM suggests that persuasion can take place at any point along the elaboration continuum, described by two different routes to persuasion: the central route, characterized by thoughtful consideration of information, and the peripheral route, which relies on superficial cues \cite{petty1986elaboration, petty2011elaboration,o2013elaboration}. The ELM offers valuable insights into how information visualization can be designed to support persuasion and decision making. Visualizations can support persuasion and decision making by facilitating both central and peripheral routes of processing. For individuals engaging in the central route, visualizations allow for deep dives into data through interactivity. For those relying on the peripheral route, well-designed visualizations with clear visual cues can trigger heuristics and emotional responses, influencing choices without extensive analysis \cite{ pandey2014persuasive,lazard2015putting, burnett2019interactive, markant2022can,lee2016effects}.
	
In summary, by offering both clear and concise visuals alongside the ability to explore details on demand, visualizations provide a valuable means to support decision making through both intuitive perception and thoughtful analysis. Unlike text-based presentations, visualizations are better at tapping into both fast, intuitive cognition and slower, analytical thinking processes from the dual-process theory, and can cater to both central and peripheral routes of processing outlined in the ELM, thus leading to quicker understanding, enhanced insight generation, and improved decision making. Recognizing the potential benefits of visual presentations, in this paper, we focus on visually explainable recommenders that augment explanations with visualizations.


\subsection{Visual Explanation} \label{sec:vis_exp}

Visualization has been increasingly used to support the understanding, diagnosis, and refinement of machine learning models \cite{liu2017towards, spinner2019explainer}. A variety of interactive graph visualization approaches were presented in the literature on XAI to explain the inner-workings of Artificial Neural Networks (ANNs) \cite{rauber2016visualizing}, Deep Neural Networks (DNNs) \cite{smilkov2017direct, wongsuphasawat2017visualizing}, Convolutional Neural Networks (CNNs) \cite{wang2020cnn, bilal2017convolutional, harley2015interactive}, Recurrent Neural Networks (RNNs) \cite{strobelt2017lstmvis, kwon2018retainvis}, and Generative Adversarial Networks (GANs) \cite{kahng2018gan}. 
All these examples highlight the need for interactive visualization during the understanding phase of XAI pipelines to improve user acceptance and enhance trust in AI systems. 
\citet{yang2020visual} investigated the effects of visual explanations on end users’ appropriate trust in machine learning. The authors concluded in their study that visual representation strongly increased users’ appropriate trust in machine learning and improved the appropriate use of the recommendations from the classifier. However, different visual explanations lead to different levels of trust and may cause inappropriate trust if an explanation is difficult to understand. 

A promising approach for XAI is visual analytics to help in bridging the gap between user knowledge and the insights XAI methods can provide \cite{spinner2019explainer, ooge2022explaining}. Visual analytics is a sub-field of information visualization, which fosters analytical reasoning through interactive visual interfaces. It uses visualization and interaction techniques to integrate human judgment in the algorithmic data analysis process. By visually exploring data and iteratively refining hypotheses, users can discover complex relations in large datasets and get insights in how algorithms work \cite{keim2008visual, thomas2006visual, cui2019visual, ooge2022explaining}. 
Many authors highlighted the benefits of visual analytics in the scope of XAI \cite{liu2017towards, endert2017state, lu2017state, hohman2018visual, chatzimparmpas2020state, spinner2019explainer, gomez2020vice, ooge2022explaining, kwon2018retainvis, choo2018visual}. For example, \citet{liu2017towards} classified visual analytics systems by whether they are intended to understand, diagnose, or refine machine learning models. Building upon this classification, \citet{spinner2019explainer} proposed a visual analytics framework for interactive and explainable machine learning that enables users to understand machine learning models, diagnose model limitations using different XAI methods, as well as refine and optimize the models. 
In terms of understanding, the authors noted that visual analytics can be applied to the interactive machine learning workflow to help different target user groups (i.e., model novices, model users, model developers) understand the model development process through tailored visual interfaces. While for a model novice, visual analytics can be used as an educational tool to explain machine learning concepts, model users and model developers would need visual analytics to understand the model’s inner logic. \citet{gomez2020vice} also presented an interactive visual analytics tool that generates counterfactual explanations (i.e., the minimal set of changes needed to flip the model’s output) to contextualize and evaluate model decisions. 
The results are displayed in a visual interface where counterfactual explanations are highlighted and interactive methods are provided for users to explore the data and model. In summary, all these works stressed the potential benefits of visual explanation in XAI, as visualization can provide an effective medium to gain insights into machine learning algorithms' outputs and make the decision-making process transparent.         

Recognizing their benefits, visualizations are also widely used to deliver explanations in recommender systems \cite{Nunes2017, zhang2020explainable}. Early works on explainable recommendation have proposed various ways of explaining recommendations to the user based on charts (e.g., bar chart, pie chart, histogram, tag cloud) and have shown that visual explanations can help to enhance transparency and users' trust, and improve the overall acceptance of a recommender system (see, e.g., \cite{Herlocker2000, Bilgic2005, Vig2009, Gedikli2014}). For example, \citet{Herlocker2000} in their seminal work of evaluating collaborative filtering-based recommender systems compared 21 different explanatory visualizations based on charts that explained the recommendation in terms of the neighborhood of similar users, and determined a ranked list of the best-performing visualizations in terms of persuasiveness and user satisfaction. The authors found in their study that the most appealing explanations were simple and conclusive methods, such as stating the neighbors’ ratings, and that more sophisticated explanations representing how the algorithm actually worked, such as a full neighbor graph, scored significantly lower. \citet{Vig2009} adopted movie tags as features to generate feature-based recommendations and explanations. To explain the recommended movie, the system uses bar charts to display the movie features and tell users why each feature is relevant to them. The authors also conducted a user study and showed that feature-based explanations can be effective even if they are in a completely different dimension (tags) than the dimension the algorithm was using for the computation (ratings).

Instead of relying on charts, recent research on visual explanation in recommender systems used images as a display style to present explanations. The survey by \citet{zhang2020explainable} provides a good entry point to image-based visually explainable recommender systems that specifically focus on deep learning techniques. To leverage the intuition of visual images, this line of research has tried to utilize item images for explainable recommendation. 
For example, \citet{LinRCRMR20} studied the task of explainable fashion recommendation (e.g., recommend a ranked list of tops for a given bottom) and proposed a convolutional neural network with a mutual attention mechanism to extract visual features of outfits, and the visual features are fed into a neural prediction network to predict the rating scores for the recommendations. During the prediction phase, the attention mechanism learns the importance of different image regions, and the importance scores indicate which parts of the image are considered when generating the recommendations. In the same fashion recommendation domain, \citet{chen2019personalized} proposed visually explainable recommendation based on personalized region-of-interest highlights. The basic idea is that for a fashion image, not all the regions are equally important for the users. Technically, the authors adopted a neural attention mechanism based on both image region-level features and user review information to discover where a user is really interested in the product image and highlight certain regions of the image as visual explanations. Research on image-based visually explainable recommendation is still at its initial stage, but this area has gained increasing attention with the continuous advancement of deep image processing techniques \cite{zhang2020explainable}. 
Image-based visual explanation over deep models is out of the scope of our survey. Rather, we concentrate primarily on visually explainable recommender systems that leverage charts commonly used in the literature on data and information visualization (e.g., bar charts, node-link diagrams) as a medium to explain their recommendations.
\subsection{Interactive Recommendation} \label{sec:intrec}
An alternate approach to increase systems' transparency explored in the literature is empowering users to control and interact with the system. Various studies from different application domains, including human-centered AI \cite{shneiderman2020bridging, shneiderman2022human}, interactive machine learning \cite{amershi2014power, dudley2018review, jiang2019recent}, and visual analytics \cite{liu2017towards, spinner2019explainer} showed that human control and interaction can also contribute to increased transparency of AI and decision making systems. As pointed out by \citet{shneiderman2022human}, interactive, visual, and exploratory user interfaces can guide users incrementally toward their goals, give them a better understanding of how the system works, and prevent their confusion and surprise that lead to the need for explanation. The benefits of human control are also well studied in the literature on interactive recommendation. We refer the interested reader to two excellent literature reviews in this area \cite{he2016interactive,jugovac2017interacting}. Interactive recommender systems provide visual and exploratory user interfaces where users can inspect the recommender process and control the system to receive better recommendations \cite{he2016interactive}. There are different ways in which users can take control over the system’s recommendations. 
Users can control the recommender system input (i.e., user profile), process (i.e., algorithm parameters), and/or output (i.e., recommendations) \cite{he2016interactive, jannach2016user}. Previous work shows that interactive recommender systems allow users to build better mental models \cite{eiband2018bringing, ngo2020exploring} and can increase transparency \cite{Tintarev2015, Tsai2017}, trust \cite{harambam2019designing, tsai2021effects}, as well as perceived effectiveness and user satisfaction \cite{jin2018effects, Pu2012, hijikata2012relation}. 

In general, human control (i.e., exploration) and explanation are complementary and both can support users in building useful mental models in recommender systems and therefore can lead to transparency \cite{Bertrand2023,Szymanski2021}.  However, they contribute to the system's transparency in different aspects \cite{tsai2021effects}. A visual exploratory user interface that allows users to control the system's input, process, and/or output cannot assure that the users build correct or complete mental models of how the system works \cite{jannach2016user}. In some cases, opening the black-box of the recommender system to users by providing explanations for system-generated recommendations has the potential to help users achieve a better understanding of the system’s functionality; thus increasing user-perceived transparency \cite{Gedikli2014, zhao2019users}. Moreover, to give feedback and interact with the system’s recommendations effectively, users need insights into the system’s reasoning, which can be achieved through explanation \cite{eiband2018bringing, ngo2020exploring, Tintarev2015}. 
In this paper, we focus on studies that combine explanation with visualization techniques to support users' understanding of and interaction with the recommendation process.  
%
%
%
\subsection{Interactive Explanation} \label{sec:intexp}
Explanation is inherently a social process \cite{miller2019explanation}. It involves the interaction between the explainer and the explainee engaging in information exchange through dialogue, visual representation, and other communication modalities \cite{hilton1990conversational}. The social nature of explanation implies that an explainable recommender system has to be interactive or even conversational \cite{liao2020questioning}. Recognizing that work on XAI  must take a human-centered approach, the HCI community has called for interdisciplinary collaboration and user-centered approaches to XAI \cite{Andjelkovic2019, wang2019designing}. With the goal of bridging the gap between XAI and HCI, research on designing and studying user interactions with XAI has emerged over the past few years \cite{abdul2018trends, liao2020questioning, cheng2019explaining, sokol2020one, krause2016interacting, kulesza2015principles, sun2022exploring, spinner2019explainer, teso2019explanatory, schramowski2020making, pfeuffer2023explanatory}. This line of research emphasizes the need to empower users to not only understand and interpret the results of AI systems through explanations but also give them the means to interact with these explanations. However, little is known about how interactive explanation should be designed and implemented in recommender systems, so that explanation aims such as scrutability, transparency, trust, and user satisfaction are met \cite{jannach2019explanations, hernandez2021effects}. Although interactive and more recently conversational recommender systems have been well studied \cite{he2016interactive, jugovac2017interacting, jannach2021survey}, there has been little work on how to incorporate interactivity features in explainable recommender system. Both in the literature and in real-world systems, there are only a few examples of recommender systems that provide interactive explanations, mainly to allow users to scrutinize the provided recommendations and correct the system’s assumptions \cite{jannach2019explanations, guesmi2021open, balog2019}, or have a conversation, i.e., an exchange of questions and answers between the user and the system, using GUI-navigation or natural language conversation \cite{hernandez2021effects}. 

A distinction is to be made here between interactive recommendation and interactive explanation. While both empower users to take control of the recommender system process, they differ in the focus and the goal of the control action.  Compared to interactive recommendation where the focus is on user interaction with the system's input, process, and/or output, interactive explanation rather focuses on interactions with the system's explanation components. Moreover, while the primary goal of interactive recommendation is to improve and personalize the recommendations, the main goal of interactive explanation is to help both the users for better understanding and the system designers for better model debugging. Our work goes beyond interactive recommendation by examining interaction possibilities to explore the reasons behind the
system’s recommendations, i.e., interactive explanation.

\subsection{Surveys} \label{sec:surveys}
Several surveys of explainable recommendation research along with taxonomies have been proposed in the literature. This research has been focused on different dimensions of explainable recommendation, such as what kind of explanations can be generated, how these have been implemented in real world systems, or the way recommendations are presented. For instance, \citet{Tintarev2007} provided a review of explanations in recommender systems by considering seven possible advantages of an explanation facility and how the ways recommendations are presented may affect explanations. 
\citet{Friedrich2011} provided a taxonomy that categorizes different explanation approaches using three dimensions, namely the reasoning model, the recommendation paradigm, and the exploited information categories. \citet{Papadimitriou2012} provided a generalized taxonomy of explanation styles for recommender systems, by categorizing them into four categories: Human, Item, Feature, and Hybrid. The authors further summarized the results of different user studies, which measure the user satisfaction for each explanation style, to conclude that Hybrid is the most effective explanation style, since it incorporates all other styles. 
\citet{Nunes2017} provided an excellent extensive review of the literature on explanations in decision-support systems. The authors discussed how explanations are generated from an algorithmic perspective as well as what kinds of information are used in this process and presented to the user. As a result, they derived a taxonomy of aspects to be considered when designing explanation in decision support systems, such as explanation objective, responsiveness, content, and presentation. More recently, \citet{zhang2020explainable} provided a two-dimensional taxonomy to classify explainable recommendation methods based on two orthogonal dimensions, namely the information source (or display
style) of the explanations and the model used to generate such explanations. The authors then provided a comprehensive
survey on model-based explainable recommendation, including factorization-based, topic modeling, (knowledge) graph-based, deep learning, knowledge-based, rule mining, and post-hoc approaches. This line of research considers the explainability of either the recommendation algorithm or the recommendation results, corresponding to model-intrinsic and model-agnostic approaches. Most of these approaches are designed to provide sentence-level explanations based on sentence templates. Based on natural
language generation models, some other approaches can automatically generate natural language textual explanations directly without using templates. Different from sentence-level textual explanations, a few other approaches have also provided visual explanations based on item images. The authors further noted that deep learning methods have been
widely adopted in model-based explainable recommender systems and that a common problem of these models is their complexity, which makes the predictions or recommendations challenging to explain.  

While there are many classifications of explanation in recommender systems, in this paper we classify visually explainable recommender systems based on four dimensions: (1) explanation aim, (2) explanation scope, (3) explanation method, and (4) explanation format. This classification is not exhaustive, and, as outlined above, there are many other, more detailed taxonomies to think about explainable recommender systems. However, for the purpose of classifying the application of visual explanation in recommender systems, our four-dimensional classification will suffice. 
This four-dimensional framework provides a concise yet effective way to systematically analyze and compare different visual explanation approaches in recommender systems from different perspectives, and captures the core aspects in the literature on explanatory visualizations in recommender systems.
Our work is different from previous survey papers	 on explainable recommendation mainly in that, to the best of our knowledge, it is the first systematic review on the topic of explanatory visualizations in recommender systems. A number of survey papers on interactive recommendation exist, considering situations in the recommendation process where users can take control (see, e.g., \cite{he2016interactive,jugovac2017interacting}).
In their review on interactive recommender systems, \citet{he2016interactive} clustered the surveyed systems based on their objectives, namely transparency, justification, controllability, and diversity. The authors discussed explanation as a means to support transparency and justification, presented different visualizations techniques adopted by the surveyed interactive recommender systems, and discussed similarities and differences between existing interactive approaches. 
\citet{jugovac2017interacting} also provides a comprehensive overview on the existing literature on user interaction aspects in recommender systems, covering existing approaches for preference elicitation, and 
result presentation, and user control over recommendations. The authors presented different visualization approaches that support user interactions and discussed explanation as one possible task of the recommendation process.
While these surveys add value to visually explainable recommendation research,
they focus mostly on information visualization
approaches supporting interaction and result presentation. 
Different from these surveys, our work mainly concentrates on information visualization to support explanation in recommender systems and provides a structured review of the research literature on explanatory visualization using a fine-grained classification scheme, based on Munzner's what-why-how visualization framework \cite{Munzner2014}. Through our systematic review, we aim to shed light on the use of explanatory visualizations in recommender systems and provide insights and guidelines which we think might be suggestive for enhancing the design of visual explanations in recommender systems. 
\section{Dimensions of Explainable Recommendation} \label{sec:dim_exp}
Explainable recommendation research covers a wide range of techniques and algorithms, and can be realized in many different ways \cite{Nunes2017}. Inspired by existing taxonomies proposed in the literature on explainable recommendation \cite{Tintarev2007,Friedrich2011,Papadimitriou2012,Nunes2017,zhang2020explainable}, in this paper, we use four dimensions to classify explainable recommendation research: (1) explanation aim, (2) explanation scope, (3) explanation method, and (4) explanation format.

\subsection{Explanation Aim} \label{sec:exp-aim}
Besides helping users understand the output and rationale of the recommender system, explanations can serve a multiplicity of aims. \citet{Tintarev2007} identified seven potential benefits that explanations may contribute to a recommender system:

\begin{itemize}
	\item \textit{Transparency}: Refers to exposing (parts of) the reasoning behind the recommendation mechanism to its users \cite{Herlocker2000}, and is defined as users’ understanding of the recommender system's inner logic \cite{Pu2012, Tintarev2007}.
	\item \textit{Scrutability}: In the context of explanations, scrutability is considered as allowing the user to tell the system it is wrong \cite{Tintarev2015} as well as correcting the system's reasoning or modifying preferences in the user model \cite{balog2019, Pu2012}. Scrutability is related to user control, as it enables the user to take action to let the system know of
mistakes in the data used for the recommendations, so future recommendations can be improved \cite{Konstan2012}.
	\item \textit{Trust}: Refers to increasing users' confidence in the system  \cite{Tintarev2007}.
	\item \textit{Effectiveness}: Can be defined as the ability of an explanation facility to help users make good
decisions \cite{Tintarev2007}. Effective explanations support users in correctly assessing if the recommended items are truly adequate for them and filtering out uninteresting items \cite{Bilgic2005, Tintarev2012}.
	\item \textit{Persuasiveness}: Refers to convincing users to try or buy certain items \cite{Tintarev2007} and to the system’s capability to nudge the user to a certain direction \cite{Nunes2017}.
	\item \textit{Efficiency}: An explanation is considered to be efficient when it helps the user make decisions faster \cite{Tintarev2007} or when it helps to reduce the cognitive effort required in the decision process \cite{Gedikli2014}. 
	\item \textit{Satisfaction}: Refers to increasing the ease of use or enjoyment. It can also be measured indirectly, measuring user loyalty \cite{Tintarev2007}. In \cite{Nunes2017} satisfaction is not considered as a single aim, but is split into ease of use, enjoyment, and usefulness.
\end{itemize}

We highlight that there are possible refinements to these aims. For example, \citet{miller2022we} differentiates between \textit{perceived trust} and \textit{demonstrated trust}. Perceived trust is measured through self-report of the user, while demonstrated trust is measured by observing the behavior of the user. The author further defines four possible outcomes when we are inciting trust in a system depending on the trustworthiness of a system and the trust a user expresses: \textit{Warranted trust}, \textit{unwarranted trust}, \textit{unwarranted distrust}, and \textit{warranted distrust}. This follows the insight, that not all trust is good trust, especially if trust is followed by accepting a proposed decision of a system. The goal must be, that a user trusts trustworthy systems while distrusting not trustworthy systems. In the recommendation domain, \citet{Gedikli2014} differentiate between \textit{objective transparency} and \textit{user-perceived transparency}. Objective transparency means that the recommender system reveals the actual mechanisms of the underlying algorithm. On the other hand, user-perceived transparency is based on the users’ subjective opinion about how well the recommender system is capable of explaining its recommendations. User-perceived transparency can be high even though the recommender system does not actually reveal the underlying recommendation algorithm \cite{Herlocker2000}. In some cases (e.g., high complexity of the algorithm) and for some users, it might be more appropriate to justify a recommendation output instead of revealing the inner workings of the recommender system \cite{Gedikli2014}. \textit{Justification} is the ability of the system to help the user understand why an item was recommended \cite{Herlocker2000}. These justifications are often more shallow and user-oriented \cite{Vig2009}. Justification is thus linked to post-hoc explanation, which aims at communicating understandable information about how an already developed model produces its predictions for any given input that may be decoupled from the initial model \cite{arrieta2020explainable}. Transparency is closely related to justification. However, transparency differs from justification in that while the former faithfully represents and exposes the reasoning about \textit{how} the recommendations are selected and \textit{how} the system works, the latter merely provides a plausible reason \textit{why} an item is recommended, that may be decoupled from the recommendation algorithm \cite{balog2019, Tintarev2015, Vig2009}.       

Other explanation aims in recommender systems include \textit{user engagement} resulting from more confidence and transparency in the recommendations \cite{Friedrich2011}, \textit{education} by allowing users to learn something from the system, and \textit{debugging} to be able to identify defects in the system and take control to make corrections \cite{Nunes2017, jannach2019explanations}.  The latter aim of debugging is closely related to scrutability \cite{lim2019these}. Additional potential aims in the context of XAI which can be borrowed in explainable recommendation include \textit{find and filter causes}, \textit{generalize and learn}, \textit{predict and control} \cite{wang2019designing, lim2019these}, as well as adoption of good practices around \textit{accountability}, \textit{ethics}, and \textit{compliance with regulations} such as the General Data Protection Regulation in Europe (GDPR) \cite{naiseh2020personalising, krebs2019tell}. To note that in the case of multiple intended explanation aims, their dependencies may be complementary (e.g., transparency may increase trust) or contradictory (e.g., persuasiveness may decrease effectiveness) \cite{Tintarev2012, Gedikli2014, Nunes2017}.
\subsection{Explanation Scope} \label{sec:exp-focus}
Explainable recommendation can consider the explainability of the recommendation input (i.e., user model), the recommendation process (i.e., algorithm), and/or the recommendation output (i.e., product). 
Explainability of the recommendation input takes the explanation to the level of user model, as opposed to that of item recommendations. That is, instead of explaining to the user why a given item was recommended, this approach provides a description that summarizes the system's understanding of the user's preferences and allows the user to scrutinize this summary and thereby directly modify his or her user model \cite{balog2019}. 
Explainable recommendation focusing on the recommendation process aims to understand how the algorithm works, and they are referred to as model-based explainable recommendation \cite{zhang2020explainable}. 
Explainability of the recommendation output focuses on the outcome, i.e., recommended items. This approach treats the recommendation process as a black box and ignores its explainability, but instead develops separate methods to explain the recommendation results produced by this black box. This approach is referred to as post-hoc explainable recommendation \cite{zhang2020explainable}. 
Compared to the explainability of the recommendation output or the recommendation process, explaining the recommendation input is under-explored in explainable recommendation research \cite{balog2019, Graus2018}.

\subsection{Explanation Method} \label{sec:exp-type}
Another way to classify explainable recommendation research is by explanation method, that is the model or strategy used for generating explanations \cite{balog2019, zhang2020explainable}. Some studies refer to the explanation method as explanation style \cite{Papadimitriou2012,Kouki2019,Tintarev2015}. However, some other researchers use the term explanation style to refer to how explanations are presented (i.e., display style) \cite{zhang2020explainable,Tsai2019}. To minimize confusion, we think that the term explanation method would be more appropriate to better describe the algorithmic mechanism used to generate explainable recommendations. In general, the underlying algorithm of a recommender engine influences to a certain degree the method used for generating explanations, e.g., a content-based recommender system produces content-based explanations \cite{Tintarev2015}. The explanation methods can broadly be categorized into four main approaches: (1) content-based, (2) collaborative-based with its two sub-categories neighborhood-based and model-based, (3) social, and (4) hybrid methods. 
\begin{itemize}
	\item \textit{Content-based} methods - also referred to as feature-based methods - attempt to match items to users based on features/attributes, such as genre, actor, director, and duration of the movies in movie recommender systems. A common paradigm for content-based explanation is to provide users with the item features that match with the user's interest profile. Because the item features/attributes are usually easily understandable to the users, it is usually intuitive to explain to the users why an item is recommended \cite{zhang2020explainable}. Content-based approaches can naturally give intuitive explanations by listing features/attributes that made an item appear in the recommendations \cite{Gemmis2015}.
	\item Instead of relying on content information, \textit{collaborative-based} methods leverage the "wisdom of the crowds" and make recommendations based on patterns of ratings or usage \cite{balog2019, Koren2015, zhang2020explainable}. Collaborative-based methods can further be broken down into neighborhood-based and model-based. \textit{Neighborhood-based} methods predict ratings by the weighted average of the ratings from like-minded users (\textit{user-based}) or ratings made by users on similar items (\textit{item-based}). The items recommended by user-based methods can be explained as ``users that are similar to you liked this item'', while item-based methods can be explained as ``the item is similar to your previously liked items''. Of the two, item-based methods often are more amenable to explaining the reasons behind the recommendations, as users are familiar with items previously preferred by them, but do not always know the other like-minded users \cite{Koren2015}. \textit{Model-based} methods capture salient characteristics of users and items by parameters learned from training data \cite{balog2019}. The most prominent model-based methods are Latent Factor Models (LFM) based on Matrix Factorization (MF) and its variants \cite{Koren2009}. The ``latent dimensions'' in these latent factor models do not possess intuitive meanings, which makes it difficult to provide intuitive recommendation explanations to the users \cite{zhang2020explainable}. To alleviate this problem, researchers provided explanations by aligning the latent dimensions with additional content information. For example, \citet{McAuley2013} aligned the latent dimensions with latent topics from Latent Dirichlet Allocation (LDA) for recommendation, \citet{Zhang2014} proposed the Explicit Factor Model (EFM) by aligning each latent dimension in matrix factorization with a particular explicit product feature extracted from textual user reviews, and \citet{Loepp2019} enhanced a latent factor model with tags users provided for the items. Deep learning models, which have recently gained much attention for explainable recommendation, also fall under the category of model-based approaches. The black box nature of deep models, however, leads to the difficulty of model explainability and using a deep model for explainable recommendation needs itself further explanation \cite{zhang2020explainable}. In general, collaborative-based methods are less intuitive to explain compared with content-based methods.
	\item \textit{Social} explanation methods leverage social friend information to provide a user with his/her social friends' interests as recommendation explanations \cite{zhang2020explainable}. For example, \citet{quijano2017make} and \citet{sharma2013social} generated a list of social friends who
also liked the recommended item as social explanations. Providing explanation with the help of social information can help to improve user trust in recommendations and explanations \cite{zhang2020explainable}, influence people’s willingness to try out an item because a trusted friend has endorsed it \cite{sharma2013social,wang2014also}, and increase user acceptance, user satisfaction, and system efficiency to help users make decisions \cite{quijano2017make}. Research results have also
shown that social explanation helps to benefit social network users in terms of different explanation aims, including transparency, scrutability, trust, persuasiveness,
effectiveness, efficiency, and satisfaction \cite{tsai2018explaining}.
	\item \textit{Hybrid} explanation methods provide explanations by aligning two or more different explanation methods \cite{Kouki2019}.
\end{itemize}

\subsection{Explanation Format} \label{sec:exp-display}
Recommendation explanations can be presented in very different display styles, which could be a relevant user or item, an image, a sentence, a chart, or a set of reasoning rules \cite{Tintarev2015, zhang2020explainable}. Generally, the format of the recommendation explanations can be classified into textual explanations and visual explanations. 

Textual explanations generate a piece of text information as recommendation explanation and can be generally classified into sentence-level and feature-level/aspect-level approaches according to how the textual explanations are displayed to users. The sentence-level methods directly present complete and semantically coherent sentences for recommendation explanations. The feature-level/aspect-level models present product features/aspects (such as color, quality, etc.) paired up with user opinion and/or sentiment on the feature/aspect. Aspects are similar to features, except that aspects are usually not directly available within an item or user profile, instead, they are extracted from e.g., the user reviews or posts \cite{zhang2020explainable, Zhang2014}.

To take advantage of the intuition of visual images, visual explanations provide the user with visualization as an explanation. The visualization can be a chart, an image (whole image or particular visual highlights in the image), or a graph, especially in social network-related application scenarios. Visual explanations can convey more information than textual ones while requiring less cognitive effort to process \cite{Ware2012}. In our survey, we focus on how visualizations can be leveraged to provide explanations in recommender systems. We use Munzner's what-why-how visualization framework \cite{Munzner2014} as a theoretical background to systematically discuss the visual explanation format approaches used in the reviewed literature. This framework is briefly explained in the next section.

\section{The What-Why-How Visualization Framework} \label{subsec:vis}
Munzner's what-why-how visualization framework \cite{Munzner2014} provides practical guidelines on how to get started on developing data visualizations, aligned with what one knows about cognition and perception. The framework breaks down visualization design according to \textit{what-why-how} questions that have \textit{data-task-idiom} answers. The what-why-how visualization framework analyzes visualization use according to three questions: \textit{what} data the user sees, \textit{why} the user intends to use a visualization tool, and \textit{how} the visual encoding and interaction idioms are constructed in terms of design choices. Each three-fold \textit{what-why-how} question has a corresponding \textit{data-task-idiom} answer. 

\subsection{What: Data Abstraction} \label{subsubsec:vis-what}
The \textit{What?} dimension refers to the abstract types of what \textit{data} can be visualized. The four basic dataset types are \textit{tables}, \textit{networks}, \textit{fields}, and \textit{geometry}; other possible collections of items include \textit{clusters}, \textit{sets}, and \textit{lists}. These datasets are made up of different combinations of the five data types: \textit{items (nodes)}, \textit{attributes}, \textit{links}, \textit{positions}, and \textit{grids}. The type of an attribute can be categorical or ordered, with a further split into ordinal and quantitative \cite{Munzner2014}.

\subsection{Why: Task Abstraction} \label{subsubsec:vis-why}
The \textit{Why?} dimension refers to tasks expressing the reason why a visualization tool is being used. A \textit{task} is defined as an \textit{\{action, target\}} pair where action is a verb defining use goals and target is a noun referring to some aspect of the data that is of interest to the user. Once a target or set of targets has been found, a low-level user goal is to query these targets. In general, visualization query actions can be at one of three scopes: \textit{identify} one target, \textit{compare} some targets, or \textit{summarize} all targets. Targets for all kinds of data are \textit{trends} (e.g., increases, decreases, peaks), \textit{outliers}, and \textit{features}, i.e., any particular task-dependent structures of interest (e.g., popularity, clusters, relationships). For one attribute, the target can be the \textit{extremes}, i.e., minimum and maximum values, or the \textit{distribution} of all values for an attribute. For multiple attributes, the target can be \textit{dependency}, \textit{correlation}, or \textit{similarity} between them. The target with network data can be \textit{topology} (i.e., structure of the network) in general or \textit{paths} in particular, and with spatial data, the target can be \textit{shape} \cite{Munzner2014}. Examples of task abstraction defined as \{action, target\} pairs include \{identify, trends\} or \{identify, outliers\} in all data; \{identify, extremes\} of one attribute; \{compare, correlation\} of two attributes; \{summarize, distribution\} of all attributes; and \{identify, paths\} in a network. The different types of actions and targets are summarized in Table \ref{tab:info-viz-why}.

\begin{table}[h]
	\caption{Task abstraction in terms of actions and targets}
	\label{tab:info-viz-why}
	\includegraphics[scale=0.5]{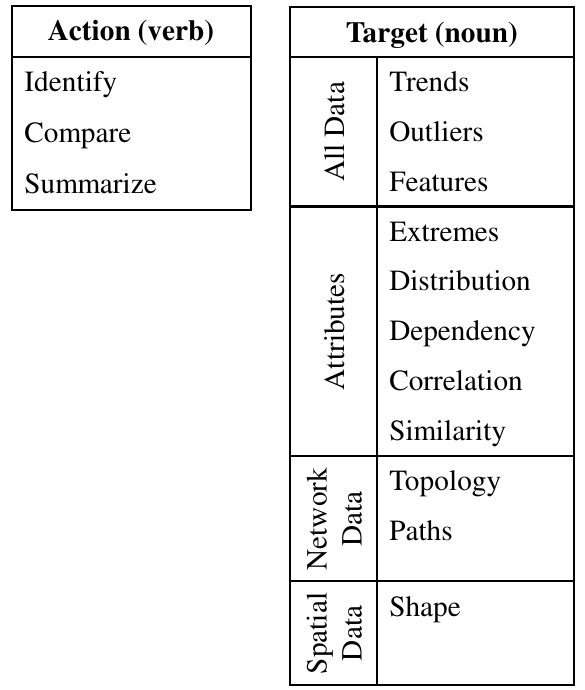}
\end{table}
\subsection{How: Idioms} \label{subsubsec:vis-how}
\begin{figure}
    \centering
   
        \includegraphics[width=0.6\textwidth]{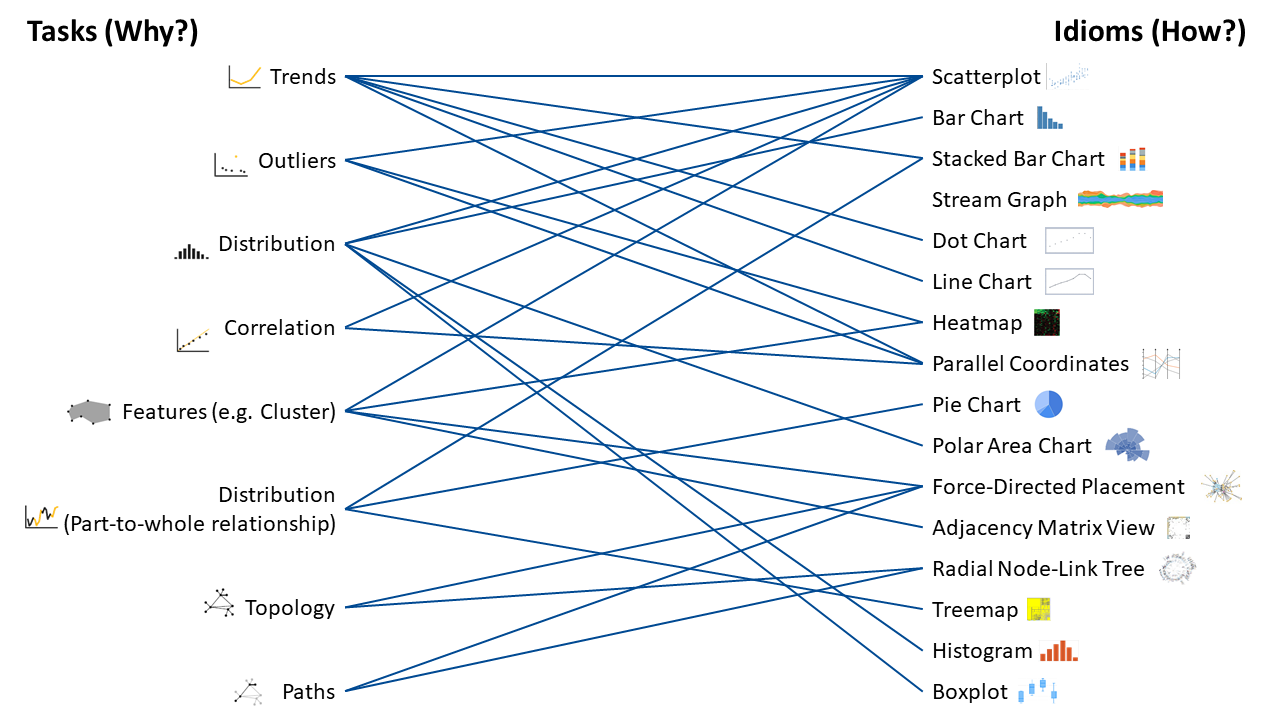}
        \caption{Data Visualization Guidelines: Mapping \textit{Why?} $\to$ \textit{How?}}
        \label{subfig:infovis-mapping-a}
 \end{figure}

\begin{figure}    
        \centering
        \includegraphics[width=0.6\textwidth]{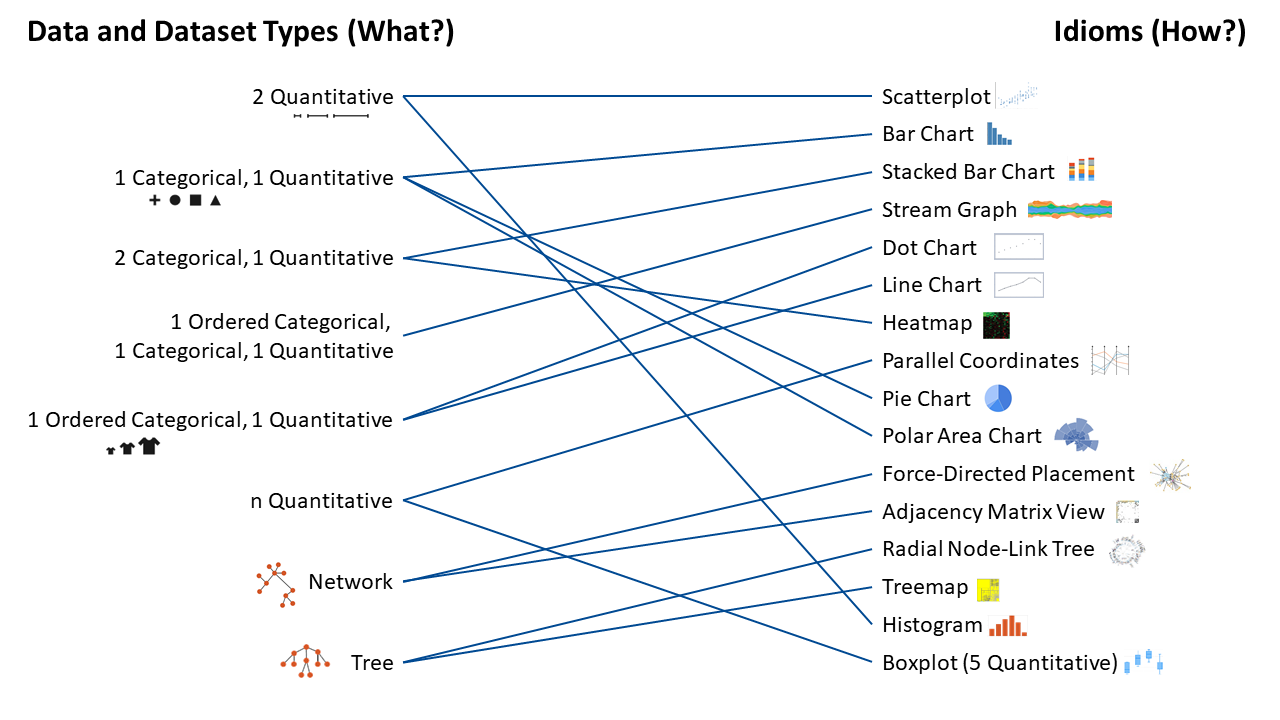}
        \caption{Data Visualization Guidelines: Mapping \textit{What?} $\to$ \textit{How?}}
        \label{subfig:infovis-mapping-b}
\end{figure}

The \textit{How?} dimension refers to \textit{how} a visualization \textit{idiom} can be constructed out of a set of design choices. These choices can be broken down into two major classes: (a) how to \textit{encode} a visualization and (b) how to \textit{interact} with a visualization. How to encode a visualization includes how to \textit{arrange} data spatially and how to \textit{map} data with all of the non-spatial visual channels, such as color, size, angle, and shape.  
Different charts (e.g., scatterplot, bar chart, line chart, heatmap, polar area chart, node-link diagram, treemap, histogram. boxplot) can be used to encode a visualization.  
Interactions include how to \textit{manipulate} a view e.g., change any aspect of the view, select elements from within the view, or navigate to change the viewpoint within the view; how to \textit{facet} data between views e.g., by juxtaposing and coordinating multiple views, partitioning data between views, or superimposing layers on top of each other; and how to \textit{reduce} the data by filtering data away, aggregating many data elements together, or embedding focus and context information together within a single view \cite{Munzner2014}.
The what-why-how visualization framework further suggests that the \textit{idioms (How?)} depend heavily on the underlying \textit{tasks (Why?)} and \textit{data (What?)} of the visualization and provides guidelines to ``what kind of idioms are more effective for what kind of tasks? (mapping \textit{Why?} $\to$ \textit{How?})'' (Figure~\ref{subfig:infovis-mapping-a}) and ``what kind of idioms is more effective for what kind of data? (mapping \textit{What?} $\to$ \textit{How?})'' (Figure~\ref{subfig:infovis-mapping-b}). For example, scatterplots and parallel coordinates are effective idioms to visualize correlation tasks. And, stacked bar charts and heatmaps are effective idioms to visualize data with two categorical and one quantitative attributes. 


\section{Systematic Review: Methodology} \label{sec:method}
A systematic review is a means of identifying, evaluating, and interpreting all available research relevant to a particular research question, topic area, or phenomenon of interest \cite{kitchenham2004}. 
In this section, we describe the steps that were taken while planning our survey on explanatory visualizations in recommender systems, following the guidelines proposed by \citet{Kitchenham2007}.
\subsection{Inclusion/Exclusion Criteria}
The literature search was conducted in June 2021.
This study included papers from journals, conferences, and book chapters, published in the English language. Out of the period of coverage or duplicated studies were excluded. A set of criteria were used to filter the primary studies retrieved from the selected databases. Table \ref{table:IE-criteria} describes the inclusion criteria (IC) and the exclusion criteria (EC).
We consider four inclusion criteria (IC) and four exclusion criteria (EC) to filter the studies to be analyzed in this review. We are interested in concrete tools (not just mockups) (IC1) that implement explainable recommendation (IC2) using visualization as a display style of explanation (IC3) and charts as a medium for visual explanation (IC4). Papers that were not written in English (EC1), that we could not access their full version (EC2), whose content is already published in more complete papers (EC3), or that solely focus on image-based visual explanations over deep models (EC4) were excluded. The latter condition excluded visually explainable recommenders like those in \cite{chen2019personalized,LinRCRMR20}. 
\begin{table}[h]
		\caption {Inclusion and exclusion criteria}
    	\label{table:IE-criteria}
	\begin{tabularx} {\textwidth} {  l	X } 
		\hline
		Inclusion Criteria (IC) & Exlusion Criteria (EC)  \\ 
		\hline
		
		\textbf{IC1:} Tool or system & 	\textbf{EC1:} Papers were not written in English  \\ 
		\textbf{IC2:} Explainable recommendation & 	\textbf{EC2:} No access to the full paper \\ 
		\textbf{IC3:} Visualization as a display style of explanation & 	\textbf{EC3:} The content of the paper was also published in another, more complete paper that is already included \\
		\textbf{IC4:} Charts as a medium for visual explanation & 	\textbf{EC4:} Focus on image-based visual explanation over deep models \\ 		
		\hline
	\end{tabularx}
\end{table}
\subsection{Search Query Construction} 
To find primary works that are relevant to our survey, we selected four main online publication databases relevant to recommender system research: ACM Digital Library, IEEE Xplore Digital Library, ScienceDirect, and Springer Link. We assumed that peer-reviewed papers published in the computer science domain are mostly stored in these selected databases. Other databases like Google Scholar and arXiv, which either automatically fetch information from various sources or allow adding non-reviewed papers by their authors, were skipped in our review. These databases would mostly provide duplicated studies. We focused on peer-reviewed work have some evidence regarding the quality of the selected studies. The search query used was \textit{(visual OR visualization OR visualisation) AND (explanation OR explainable) AND (recommender OR recommendation) AND (tool OR system)}. The query was run in the four academic databases, resulting in 235 papers (excluding duplicates). It was customized to each of the target databases since each one has its specific syntax.

\subsection{Primary Studies Selection}
 \begin{table}[h]
 	\centering
 	\caption {Search results for each database}
 	\label{table:query-rslt}
 	\begin{tabular}  {  l	r } 
 		\hline
 		Database & number of papers  \\ 
 		\hline
 		
 		ACM  Digital Library & 	44  \\ 
    	IEEE Xplore  Digital Library & 278 \\ 
    	ScienceDirect & 84 \\ 		
 		Springer Link & 34 \\ 		
 		
 		\hline
 		
 		\textbf{Total} & 440 \\
 		\hline
 	\end{tabular}
 \end{table}
 
After executing our query on the four selected databases we obtained 440 papers. Table \ref{table:query-rslt} provides statistical details for each database. The relevant primary tools were then selected using a three-step procedure. In the first step, the third author excluded duplicates resulting in 235 papers. In the second step, the remaining papers were randomly distributed among the three authors to ensure that each paper was reviewed by at least two researchers. Each author analyzed the title and abstract of each of the assigned papers and looked for screenshots of the proposed system in the paper. If the title or abstract provided any sort of indication that the paper describes the concrete implementation of a specific visual explanation in a recommender system (not just a study based on mockups or sample visualization interfaces), the paper was selected to be subsequently analyzed in detail. In the third step, the full text of each paper was retrieved, if available, and analyzed. Then, each author re-evaluated whether their assigned papers truly met the above criteria. It is worth noting that in the last two steps of the review, conflicting viewpoints or confusing papers were discussed to reach a consensus by all three authors. At the end, we ended up with a total of 33 primary tools that were included in the final analysis.
\section{Results} \label{sec:result}
In this section, we summarize the insights we obtained by analyzing the 33 selected visually explainable recommendation tools, based on the four dimensions introduced in Section \ref{sec:exp}.

\begin{table}[h]
	\caption{Summary of the survey results - explanation aim, scope, and method}
	\label{table:survey-aim-focus-type}
	\setlength\tabcolsep{2.3pt}
	\begin{tabular}{|l|p{1.0mm}|p{1.0mm}|p{1.0mm}|p{1.0mm}|p{1.0mm}|p{1.0mm}|p{1.0mm}|p{1.0mm}|p{1.0mm}|p{1.0mm}|p{1.0mm}|p{1.0mm}|p{1.0mm}|p{1.0mm}|p{1.0mm}|p{1.0mm}|p{1.0mm}|}
		\hline
		& 
		\multicolumn{8}{c|}{\textbf{Aim}} & 
		\multicolumn{3}{c|}{\textbf{Scope}} & 
		\multicolumn{6}{c|}{\textbf{Method}} \\ \cline{2-18}
		
		\multicolumn{1}{|c|}{\textbf{Tools}} & 
		\multicolumn{1}{c|}{\rotatebox{270}{\scriptsize{Transparency}}} & 
		\multicolumn{1}{c|}{\rotatebox{270}{\scriptsize{Justification}}} & 
		\multicolumn{1}{c|}{\rotatebox{270}{\scriptsize{Scrutability}}} & 
		\multicolumn{1}{c|}{\rotatebox{270}{\scriptsize{Trust}}} &
		\multicolumn{1}{c|}{\rotatebox{270}{\scriptsize{Effectiveness}}} & 
		\multicolumn{1}{c|}{\rotatebox{270}{\scriptsize{Persuasiveness}}} & 
		\multicolumn{1}{c|}{\rotatebox{270}{\scriptsize{Efficiency}}} & 
		\multicolumn{1}{c|}{\rotatebox{270}{\scriptsize{Satisfaction}}} &
		
		\multicolumn{1}{c|}{\rotatebox{270}{\scriptsize{Input}}} & 
		\multicolumn{1}{c|}{\rotatebox{270}{\scriptsize{Process}}} & 
		\multicolumn{1}{c|}{\rotatebox{270}{\scriptsize{Output}}} & 
		
		\multicolumn{1}{c|}{\rotatebox{270}{\scriptsize{Content-based}}} & 
		\multicolumn{1}{c|}{\rotatebox{270}{\scriptsize{CF-Neigh-User }}} & 
		\multicolumn{1}{c|}{\rotatebox{270}{\scriptsize{CF-Neigh-Item }}} & 
		\multicolumn{1}{c|}{\rotatebox{270}{\scriptsize{CF-Model}}} & 
		\multicolumn{1}{c|}{\rotatebox{270}{\scriptsize{Social}}} &
		\multicolumn{1}{c|}{\rotatebox{270}{\scriptsize{Hybrid}}} \\ \hline
		
		\tiny{\textbf{Relevance Tuner+ }(\citet{Tsai2019})} &  & \cellcolor{green!25}+ &  &  &  & \cellcolor{green!25}+ &  & \cellcolor{green!25}+ &  &  & + & + &  &  &  & + &  \\ \hline
		\tiny{\textbf{HyPER} (\citet{Kouki2019})} &  & + &  &  &  & \cellcolor{red!25}+ &  & + &  &  & + & + & + & + &  & + & + \\ \hline
		\tiny{\textbf{Tagsplanation} (\citet{Vig2009})} &  & \cellcolor{green!25}+ &  &  & \cellcolor{green!25}+ &  &  &  &  &  & + & + &  &  &  &  &  \\ \hline
		\tiny{\textbf{PeerChooser} (\citet{ODonovan2008})} & + &  &  & + &  &  &  &  &  & + & + &  & + &  &  &  &  \\ \hline
		\tiny{\textbf{TalkExplorer} (\citet{Verbert2013})} &  & + &  & + & \cellcolor{green!25}+ &  &  &  &  &  & + & + &  &  &  &  &  \\ \hline
		\tiny{\textbf{SetFusion} (\citet{Parra2014})} & + &  &  &  &  &  &  &  &  & + & + & + &  &  &  &  &  \\ \hline
		\tiny{\textbf{TasteWeights} (\citet{Bostandjiev2012})} & + &  &  &  &  &  &  & \cellcolor{green!25}+ &  & + & + & + &  &  &  & + & + \\ \hline
		\tiny{\textbf{Moodplay} (\citet{Andjelkovic2019})} &  & + &  &  &  &  &  & \cellcolor{green!25}+ &  &  & + & + &  &  &  &  &  \\ \hline
		\tiny{\textbf{SmallWorld} (\citet{Gretarsson2010})} & + &  &  &  &  &  &  & \cellcolor{green!25}+ &  & + & + &  & + &  &  & + & + \\ \hline
		\tiny{\textbf{LinkedVis} (\citet{Bostandjiev2013})} & + &  &  &  & + &  &  & +\cellcolor{green!25} &  & + & + & + &  &  &  & + & + \\ \hline
		\tiny{(\citet{Schaffer2015})} & + &  &  & \cellcolor{green!25}+ &  &  &  & \cellcolor{green!25}+ &  & + & + &  & + &  &  &  &  \\ \hline
		\tiny{(\citet{Vlachos2012})} &  & + &  &  &  &  &  &  &  &  & + & + &  &  &  &  &  \\ \hline
		\tiny{\textbf{PARIS-ad} (\citet{Jin2016})} & \cellcolor{green!25}+ &  & + &  &  & + &  &  & + & + & + & + &  &  &  &  &  \\ \hline
		\tiny{(\citet{Bakalov2013})} &  & + & + & \cellcolor{green!25}+ & \cellcolor{green!25}+ &  &  & \cellcolor{green!25}+ & + &  & + & + &  &  &  &  &  \\ \hline
		\tiny{(\citet{Kangasraeaesioe2015})} &  & + &  &  & \cellcolor{green!25}+ &  &  & \cellcolor{green!25}+ &  &  & + & + &  &  &  &  &  \\ \hline
		\tiny{\textbf{System U} (\citet{Badenes2014})} &  & + &  &  &  &  &  &  & + &  &  & + &  &  &  &  &  \\ \hline
		\tiny{\textbf{UniWalk} (\citet{Park2017})} &  & + &  &  &  & + &  &  &  &  & + &  & + & + &  & + & + \\ \hline
		\tiny{(\citet{Millecamp2019})} &  & + & + & \cellcolor{green!25}+ & \cellcolor{green!25}+ &  &  & \cellcolor{green!25}+ & + &  & + & + &  &  &  &  &  \\ \hline
		\tiny{\textbf{RelExplorer} (\citet{Tsai2017})} &  & + &  &  & \cellcolor{green!25}+ &  &  & + &  &  & + & + &  &  &  & + &  \\ \hline
		\tiny{\textbf{Pharos} (\citet{Zhao2011})} &  & + &  & + & \cellcolor{green!25}+ &  &  &  &  &  & + & + &  &  &  &  &  \\ \hline
		\tiny{(\citet{Zhang2014})} &  & + &  &  &  & \cellcolor{green!25}+ &  &  &  &  & + & + &  &  &  &  &  \\ \hline
		\tiny{\textbf{LIBRA} (\citet{Bilgic2005})} &  & + &  &  & \cellcolor{yellow!25}+ &  &  &  &  &  & + &  & + &  &  &  &  \\ \hline
		\tiny{\textbf{PeerFinder} (\citet{Du2018})} &  & + &  &  & \cellcolor{green!25}+ &  &  &  & + &  & + & + &  &  &  &  &  \\ \hline
		\tiny{(\citet{Dominguez2018})} &  & + &  & \cellcolor{green!25}+ &  &  &  & \cellcolor{green!25}+ &  &  & + & + &  &  &  &  &  \\ \hline
		\tiny{(\citet{Mueller2019})} &  & + & + & + &  &  &  & + & + &  & + & + &  &  &  &  &  \\ \hline
		\tiny{\textbf{Mastery Grids} (\citet{Barria-Pineda2019})}  & + &  &  &  &  &  &  &  & + & + & + & + &  &  &  &  &  \\ \hline
		\tiny{\textbf{ComBub} (\citet{Jin2018})} &  & + &  &  &  &  &  & \cellcolor{green!25}+ &  &  & + & + &  &  &  &  &  \\ \hline
		\tiny{\textbf{CourseQ} (\citet{ma2021courseq})} & \cellcolor{green!25}+ &  &  & \cellcolor{green!25}+ & \cellcolor{green!25}+ &  &  & \cellcolor{green!25}+ &  & + & + &  &  &  & +  &  &  \\ \hline
		\tiny{\textbf{BrainHood} (\citet{tsiakas2020brainhood})} & + &  &  &  &  & + &  &  & + &  & + & + &  &  &  & + & + \\ \hline
		\tiny{\textbf{Rankfomsets} (\citet{bansal2020recommending})} & + &  & + &  &  &  &  &  & + &  & + &  &  &  & + &  &  \\ \hline
		\tiny{\textbf{IW Browser} (\citet{narayanan2008towards})} & + &  &  & + &  &  &  &  &  & + & + & + &  &  &  &  &  \\ \hline
		\tiny{\textbf{uRank} (\citet{sciascio2017supporting})} & + &  & + &  &  &  &  &  & + &  & + & + &  &  &  &  &  \\ \hline
		\tiny{(\citet{alshammari2019mining})} &  & + &  &  & \cellcolor{green!25}+ &  &  & \cellcolor{green!25}+ &  & + & + &  &  &  & + &  &  \\ \hline
		
	\end{tabular}
	\caption*{Evaluation results: \crule[green!25]{0.2cm}{0.2cm} positive effect \crule[red!25]{0.2cm}{0.2cm} negative effect \crule[yellow!25]{0.2cm}{0.2cm} mixed effect}
\end{table}

\subsection{Explanation Aim}
We used the list of potential explanation aims presented in Section \ref{sec:exp-aim} to categorize the different tools. We noted that the explanation aims were not mentioned explicitly in most of the papers presenting the tools analyzed in our review. Therefore, we tried to identify the aims based on the description or if available the evaluation of the tool.  We did not find studies analyzed in our review that focus on additional aims (e.g., education, debugging) beyond the initial list of possible explanation aims proposed by \citet{Tintarev2007}. We further considered \textit{justification} as an explanation aim when the tool provides an abstract description of how recommendations are generated rather than the reasoning behind the recommendation algorithm to explain how the system works.

\begin{figure}[h]
	\centering
		\includegraphics[width=0.5\textwidth]{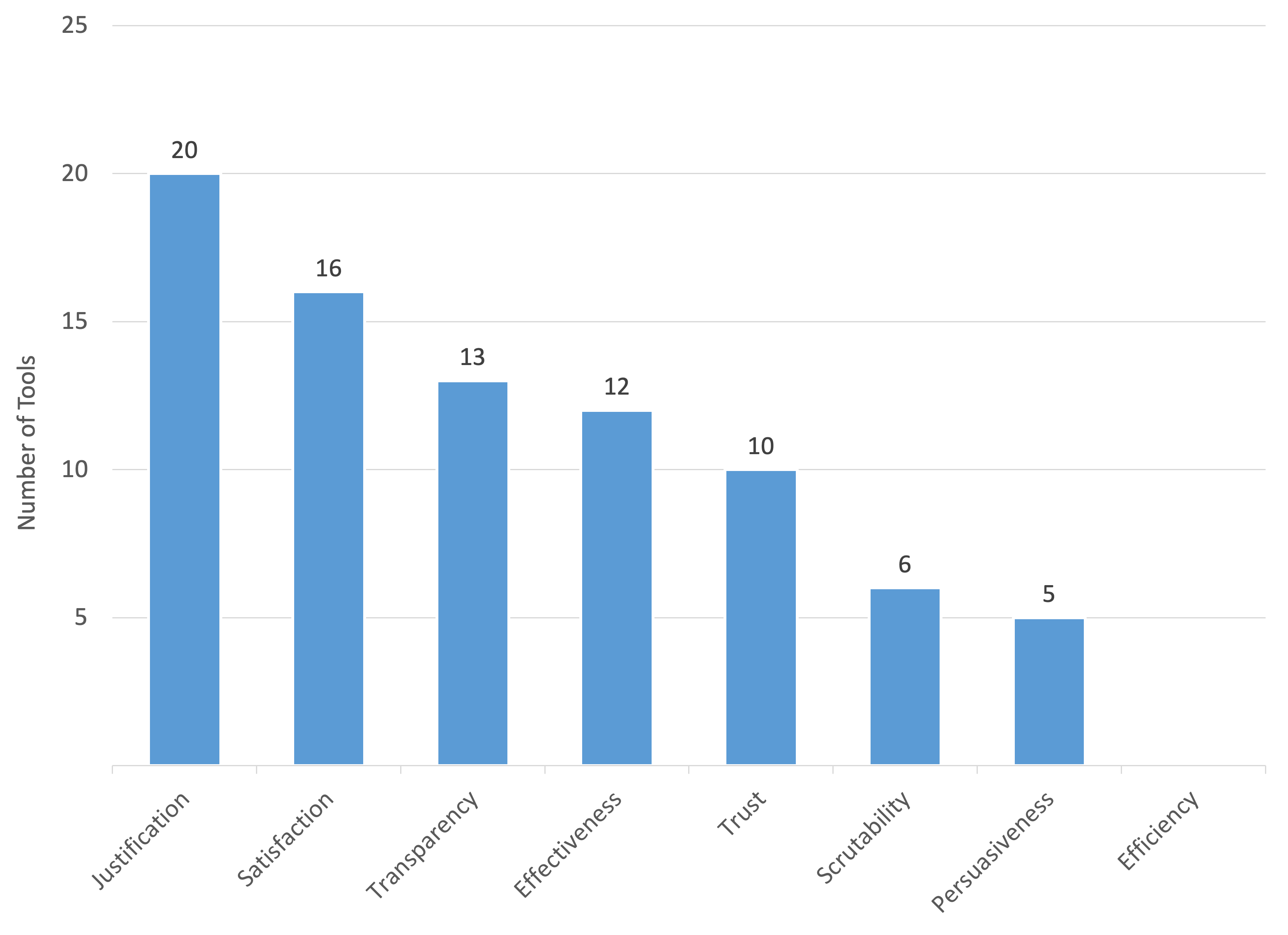}
		\caption{Explanation aim}
		\label{fig:survey-aim}
	\end{figure}

The majority of the tools analyzed in our review focused on more than one aim. 
The maximum number of aims (five) was identified for the tools in \cite{Bakalov2013, Millecamp2019}. 
As shown in the explanation aim part in Table \ref{table:survey-aim-focus-type} and Figure \ref{fig:survey-aim}, the most common explanation aim is to provide \textit{justification}, which is the basic aim that any explanation needs to support. Thirteen tools focused on \textit{transparency} by providing insights into parts of the inner logic of the recommendation algorithm. 
While \textit{satisfaction} and \textit{effectiveness} were the focus of a considerable number of tools, \textit{persuasiveness} and \textit{scrutability} were under-explored in the analyzed tools, but are receiving more attention in recent years. None of the tools aimed at \textit{efficiency}, which supports prior findings \cite{Gedikli2014} that for users making good decisions (i.e., effectiveness) is more important than making decisions faster.

The authors of the surveyed tools reported user studies evaluating the effects of visual explanations on different aims, with contradicting results. It was found that providing visual explanation generally has positive effects on transparency \cite{Jin2016, ma2021courseq}, justification \cite{Tsai2019, Vig2009}, trust \cite{Schaffer2015, Bakalov2013, Millecamp2019, Dominguez2018, ma2021courseq}, effectiveness \cite{Vig2009, Verbert2013, Bakalov2013, Kangasraeaesioe2015, Millecamp2019, Tsai2017, Zhao2011, Bilgic2005, Du2018, ma2021courseq, alshammari2019mining}, persuasiveness \cite{Tsai2019, Zhang2014}, and satisfaction \cite{Tsai2019, Bostandjiev2012, Andjelkovic2019, Gretarsson2010, Bostandjiev2013, Schaffer2015, Bakalov2013, Kangasraeaesioe2015, Millecamp2019, Dominguez2018, Jin2018, ma2021courseq, alshammari2019mining}. On the other hand, two studies reported negative effects on persuasiveness \cite{Kouki2019} and effectiveness \cite{Bilgic2005}. \citet{Kouki2019} found significant evidence that all visual explanation  formats they provided (i.e., Venn diagrams and cluster dendrograms) were perceived as less persuasive than text explanations. \citet{Bilgic2005} found that neighborhood style explanation for collaborative filtering systems using bar charts actually causes users to overestimate the quality of an item. 

Only one study \cite{Bilgic2005} explored the effects of different visual explanations on one intended aim. Focusing on effectiveness as an explanation aim, \citet{Bilgic2005} found that while the neighborhood style explanation based on bar charts caused users to overestimate the quality of an item, keyword-style explanations, and influence-style explanations, based on simple lists of keywords and ratings, respectively, were found to be significantly more effective at enabling accurate assessments. 
Moreover, only the study in \cite{Tsai2019} investigated the effects of different visual explanations on multiple aims in parallel. This study showed evidence that different visual explanations result in different levels of perceptions of the explainable recommender systems, in terms of different intended explanation aims. Concretely, \citet{Tsai2019} found that, for better justification, the explanation format of word cloud, graph, and map received higher scores. For better persuasiveness, the explanation format of the graph outperformed the other visualizations. And, for better satisfaction, the explanation formats of word cloud and graph were preferred by the participants after experiencing the system.

Further, few studies pointed out that there are potential complementary relationships between different intended explanation aims. For example, justification and transparency are considered as essential to develop trust toward the recommender system \cite{Verbert2013, Schaffer2015, Bakalov2013, Millecamp2019, Dominguez2018, ma2021courseq}. However, these complementary relationships were not systematically evaluated. And, none of the studies reported contradictory dependencies between multiple intended explanation aims.  


%
\subsection{Explanation Scope}
\begin{figure}[h]
		\includegraphics[width=0.5\textwidth]{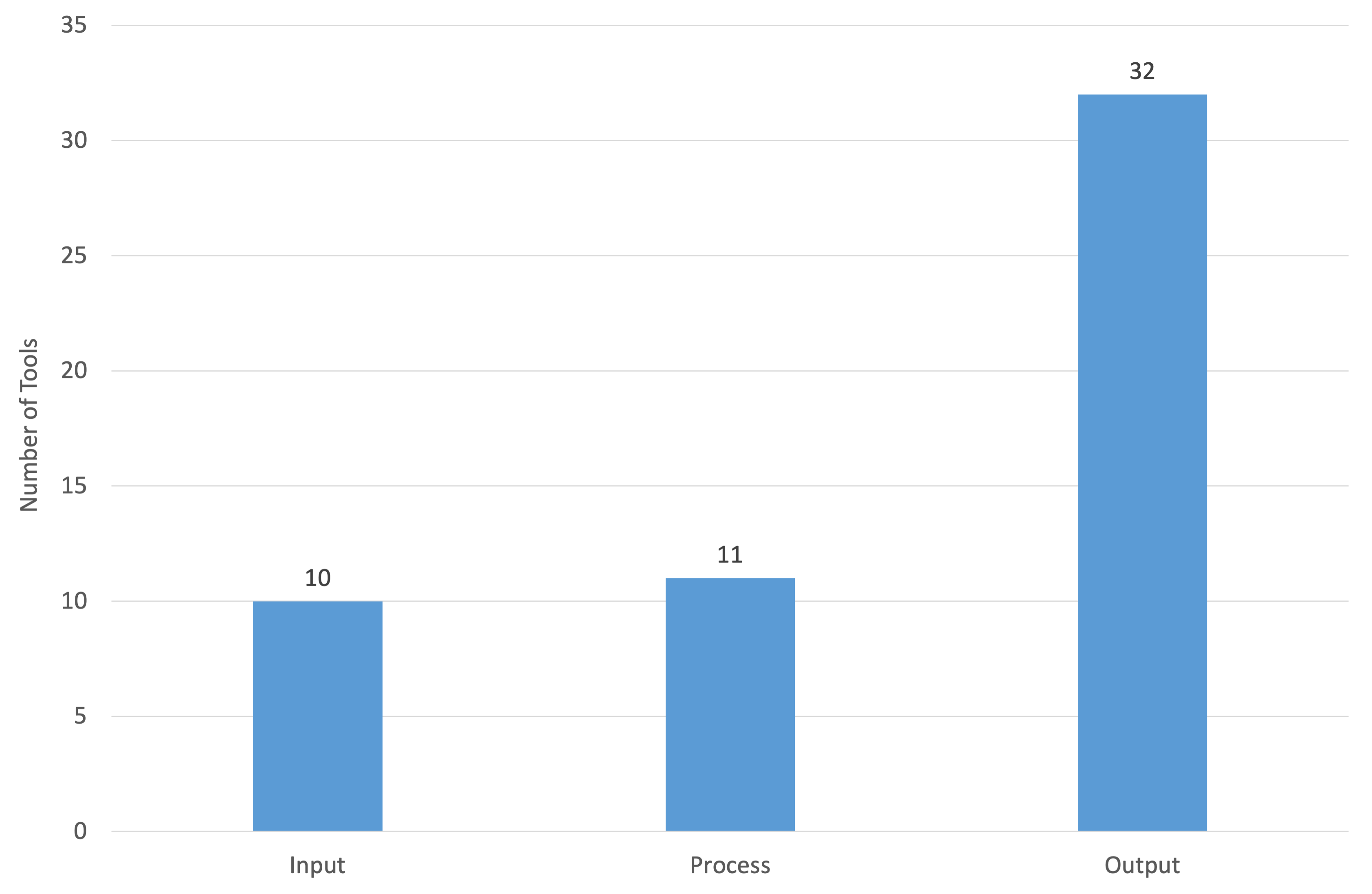}
		\caption{Explanation scope}
		\label{fig:survey-focus}
	\end{figure}

In this dimension, the surveyed tools are classified based on the part of a recommender system they are trying to explain, i.e., the recommendation \textit{input}, \textit{process}, and/or \textit{output}. As can be seen in the explanation scope part of Table \ref{table:survey-aim-focus-type} and Figure \ref{fig:survey-focus}, all the analyzed tools except `System U' \cite{Badenes2014} provide explanations of the recommendation output. For example, `HyPER' explains the recommended music artist based on the artist's popularity, similarity with the artists in the user profile, and artists liked by similar users \cite{Kouki2019}. `Moodplay' also recommends music artists but explains them based on the mood (e.g., sad, joyful, serious) of the songs by the artists included as input for the recommendation \cite{Andjelkovic2019}. The data from the Conference Navigator 3 (CN3) system is used by `Relevance Tuner+' \cite{Tsai2019} and `RelExplorer' \cite{Tsai2017} to explain the recommended conference talks and possible scholars to collaborate with based on social and academic similarity. The tools \cite{Schaffer2015}, \cite{Vlachos2012},  `Tagsplanation' \cite{Vig2009}, and \cite{alshammari2019mining} recommend movies and provide visual explanations based on community tags, similar users, similar movie content (bag-of-words), and similar interests (semantic attributes), respectively. 

To make the recommendation algorithm transparent, 11 tools focused on providing visual explanations for their recommendation process. `PARIS-ad' \cite{Jin2016}, for instance, provides a simple flow chart as an explanation that exposes the process of recommending an advertisement. `SmallWorld' \cite{Gretarsson2010} uses complex network visualization based on five layers to visualize the inner logic of the recommendation process to explain the connection between the active user and recommended items. The tools `LinkedVis' \cite{Bostandjiev2013}, `TasteWeights' \cite{Bostandjiev2012}, and `PeerChooser' \cite{ODonovan2008}  try to explain the recommendation process by highlighting the relationships between user profile attributes (input), social connections, and the recommended items (output). `Mastery Grids' \cite{Barria-Pineda2019} highlights the Java programming concepts and their weightage which has contributed to the selected recommended activity (e.g., quiz, example, coding) for students. 'IW browser' \cite{narayanan2008towards} generates intuitive explanations for career counseling using multiple display styles, including a graphical one using a node-link diagram to describe the inference steps used to derive the explanation.

The explanation of the inputs of the recommendation is a relatively new explanation scope and is supported by ten tools. `System U' \cite{Badenes2014} explains the input by visually exposing the user model, which consists of Big Five personality characteristics, fundamental needs, and human values. To make students understand why a certain learning activity is recommended to them, `Mastery Grids' \cite{Barria-Pineda2019} highlights the concepts related to the recommended activity based on fine-grained open learner models. The exposed user model in `PeerFinder' \cite{Du2018} consists of different university student features (e.g., gender, age, program) and it is used to recommend similar peers. 'BrainHood' \cite{tsiakas2020brainhood} uses different visualizations of open learner models to help children monitor their performance. In addition to exposing the user model, the tools in \cite{Bakalov2013}, \cite{Mueller2019}, \cite{Jin2016}, \cite{Millecamp2019}, \cite{bansal2020recommending}, and \cite{sciascio2017supporting} allow users to scrutinize and modify their models to adjust the recommendation results according to their needs and preferences. However, none of these tools provide an explanation of how a user model was generated.
%
\subsection{Explanation Method}
\begin{figure}[h]
		\includegraphics[width=0.5\textwidth]{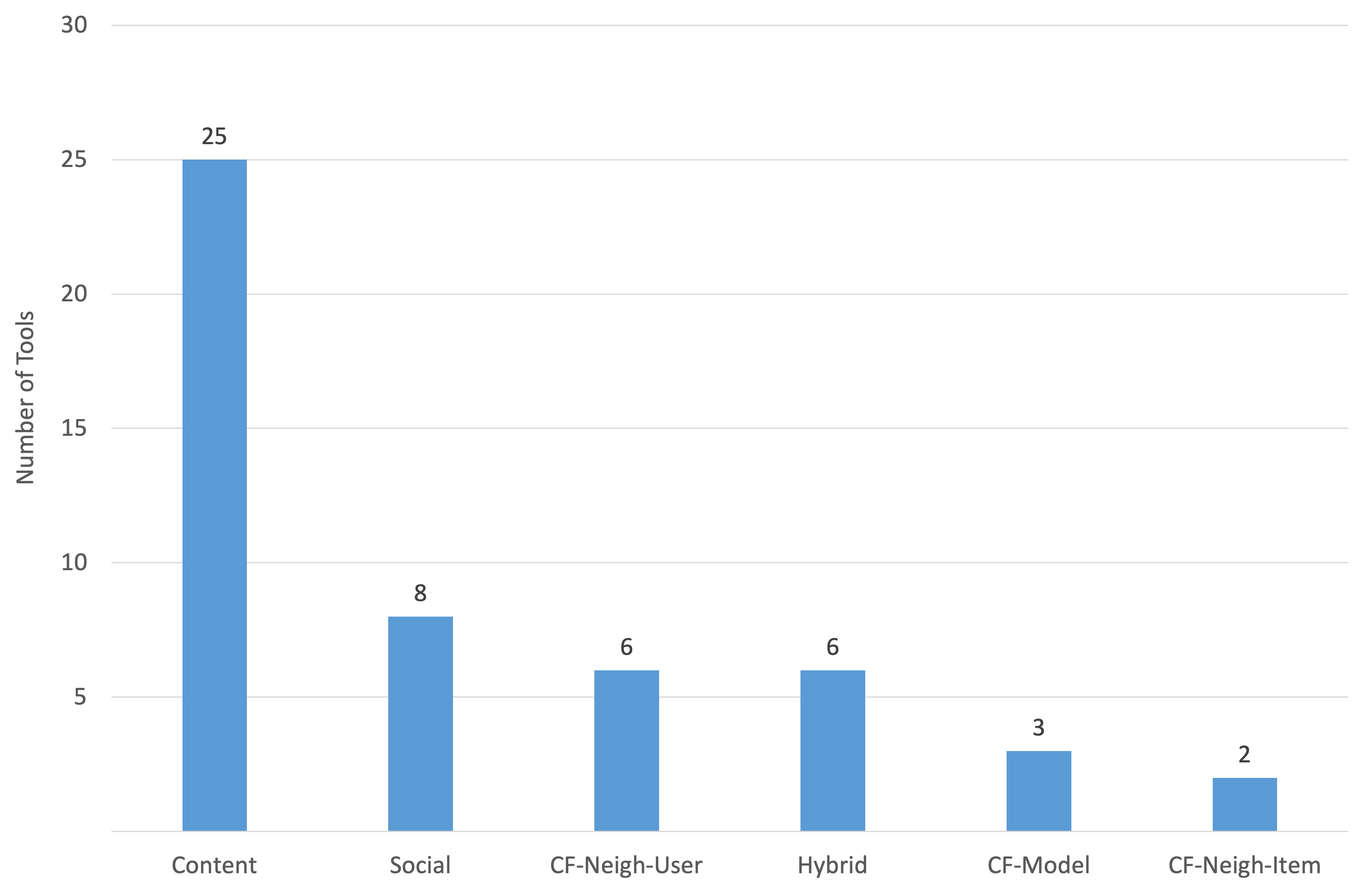}
		\caption{Explanation method}
		\label{fig:survey-type}
	\end{figure}

Explanation method is the third dimension in this analysis, which focuses on how the explanation is generated. We noted that most of the reviewed works do not provide details about the explanation generation process. However, often the underlying recommendation algorithm is discussed. We also observed that in most cases the presented explanation is closely tied to the underlying recommendation algorithm, which confirms the findings of other researchers that in general, the used recommendation algorithm controls what kind of explanations could be generated \cite{Tintarev2015,Nunes2017}. To identify the explanation method in each of the reviewed tools, we analyzed the provided visualization of the explanation together with the details of the underlying recommendation algorithm. Some tools use different explanation methods. We considered that a tool follows a hybrid explanation approach if multiple explanation methods are combined within a single visualization. The outcome of this analysis is shown in the explanation method part of Table \ref{table:survey-aim-focus-type} and Figure \ref{fig:survey-type}. 

The majority of the examined tools adopted a \textit{content-based} approach for an explanation. For example, `ComBub' explains the diversity of the recommended songs by visualizing their selected audio features (e.g., valence, liveness, genre) \cite{Jin2018} and the tool in \cite{Dominguez2018} explains the recommended images based on their attractiveness visual features, such as brightness, contrast, and colorfulness. This is not surprising given the fact that content-based approaches can give intuitive explanations to the users why an item is recommended by listing features/attributes that contribute to the recommendation (see Section \ref{sec:exp-type}). Only six tools used the \textit{collaborative-based} methods, mainly the \textit{user-based} approach. For instance, `SmallWorld' recommends items to an active user based on a collaborative analysis of their preference data from his/her direct peer group on a social network \cite{Gretarsson2010} and the tool in \cite{Schaffer2015} recommends movies to a user based on their previous ratings and similar users. The \textit{model-based} approach is not frequently used in the tools that we investigated. This might be due to the complexity of this approach, which makes it difficult to provide intuitive recommendation explanations to the users, as argued by \citet{zhang2020explainable}. In our survey, this approach was adopted in only three studies
\cite{alshammari2019mining, bansal2020recommending, ma2021courseq}. For instance, \citet{alshammari2019mining} uses semantic knowledge graphs (KG) that correlate the user with the item’s semantic attributes based on the number of interactions between them in the user’s history. The semantic KGs are used in the latent spaces to build the final model and to generate justifications for the recommendations.
Eight tools adopted the \textit{social} explanation method, out of which five tools used the \textit{hybrid} approach to combine the \textit{social} explanation method with one or more other methods. 
\subsection{Explanation Format}
The fourth dimension in our analysis deals with how the explanations are displayed to the user. In this work, we focus on visual explanations and use Munzner's what-why-how visualization framework introduced in Section \ref{subsec:vis} to systematically compare the visual explanation format approaches used in the reviewed tools.

\begin{table} [h]
	\caption{Summary of the survey results - explanation format (What / Why)}
	\label{table:survey-what-why}
	\setlength\tabcolsep{4pt}
	\begin{tabular}{|l|p{1.0mm}|p{1.0mm}|p{1.0mm}|p{1.0mm}|p{1.0mm}|p{1.0mm}|p{1.0mm}|p{1.0mm}|p{1.0mm}|p{1.0mm}|p{1.0mm}|}
		\hline
		& 
		\multicolumn{4}{c|}{\textbf{What}} & 
		\multicolumn{7}{c|}{\textbf{Why}} \\ \cline{2-12}

		& 
		\multicolumn{4}{c|}{\textbf{Dataset}} & 
		\multicolumn{7}{c|}{\textbf{Task}} \\ \cline{2-12}
		
		\multicolumn{1}{|c|}{\textbf{Tools}} & 
		\multicolumn{1}{c|}{\rotatebox{270}{\scriptsize{Tables}}} & 
		\multicolumn{1}{c|}{\rotatebox{270}{\scriptsize{Networks}}} & 
		\multicolumn{1}{c|}{\rotatebox{270}{\scriptsize{Sets}}} & 
		\multicolumn{1}{c|}{\rotatebox{270}{\scriptsize{Geometry}}} & 
		
		\multicolumn{1}{c|}{\rotatebox{270}{\{\scriptsize{Identify, Paths\}}}} & 
		\multicolumn{1}{c|}{\rotatebox{270}{\{\scriptsize{Identify, Topology\}}}} & 
		\multicolumn{1}{c|}{\rotatebox{270}{\{\scriptsize{Compare, Similarity\}}}} & 
		\multicolumn{1}{c|}{\rotatebox{270}{\{\scriptsize{Summarize, Similarity\}}}} & 
		\multicolumn{1}{c|}{\rotatebox{270}{\{\scriptsize{Identify, Features\}}}} & 
		\multicolumn{1}{c|}{\rotatebox{270}{\{\scriptsize{Summarize, Features\}}}} & 
		\multicolumn{1}{c|}{\rotatebox{270}{\{\scriptsize{Identify, Distribution\}}}} \\ \hline

		\scriptsize{\textbf{Relevance Tuner+} (\citet{Tsai2019})} & + & + & + & + & + &  & + &  &  &  &  \\ \hline 
		\scriptsize{\textbf{HyPER} (\citet{Kouki2019})} & + &  & + &  &  &  & + & + &  &  &   \\ \hline 
		\scriptsize{\textbf{Tagsplanation} (\citet{Vig2009})} & + &  &  &  &  &  & + &  &  &  & +  \\ \hline  
		\scriptsize{\textbf{PeerChooser} (\citet{ODonovan2008})} & + &  &  &  & + & + &  &  &  &  &    \\ \hline 
		\scriptsize{\textbf{TalkExplorer} (\citet{Verbert2013})} &  & + &  &  & + & + &  &  &  &  &    \\ \hline 
		\scriptsize{\textbf{SetFusion} (\citet{Parra2014})} &  &  & + &  &  &  &  & + &  &  &   \\ \hline 
		\scriptsize{\textbf{TasteWeights} (\citet{Bostandjiev2012})} &  & + &  &  & + &  &  &  &  &  &   \\ \hline  
		\scriptsize{\textbf{Moodplay} (\citet{Andjelkovic2019})} & + &  &  &  &  &  &  & + & + &  &    \\ \hline 
		\scriptsize{\textbf{SmallWorld} (\citet{Gretarsson2010})} &  & + &  &  & + & + &  &  &  &  &   \\ \hline 
		\scriptsize{\textbf{LinkedVis} (\citet{Bostandjiev2013})} &  & + &  &  & + &  &  &  &  &  &   \\ \hline 
		\scriptsize{(\citet{Schaffer2015})} & + &  &  &  & + &  &  &  &  &  &   \\ \hline  
		\scriptsize{(\citet{Vlachos2012})} &  & + &  &  & + &  & + &  &  &  &    \\ \hline 
		\scriptsize{\textbf{PARIS-ad} (\citet{Jin2016})} & + &  &  &  & + &  &  &  &  &  &   \\ \hline 
		\scriptsize{(\citet{Bakalov2013})} & + &  &  &  &  &  &  &  & + & + &    \\ \hline 
		\scriptsize{(\citet{Kangasraeaesioe2015})} & + &  &  &  &  &  & + &  &  &  &  \\ \hline 
		\scriptsize{\textbf{System U} (\citet{Badenes2014})} & + &  &  &  &  &  &  &  &  & + & +  \\ \hline 
		\scriptsize{\textbf{UniWalk} (\citet{Park2017})} & + &  &  &  &  &  & + &  &  &  &  \\ \hline 
		\scriptsize{(\citet{Millecamp2019})} & + &  &  &  &  &  & + &  &  &  &   \\ \hline 
		\scriptsize{\textbf{RelExplorer} (\citet{Tsai2017})} &  & + & + &  & + &  & + &  &  &  &  \\ \hline 
		\scriptsize{\textbf{Pharos} (\citet{Zhao2011})} &  &  & + &  &  &  &  & + &  &  &   \\ \hline 
		\scriptsize{(\citet{Zhang2014})} &  &  & + &  &  &  &  &  & + & + &   \\ \hline 		
		\scriptsize{\textbf{LIBRA} (\citet{Bilgic2005})} & + &  &  &  &  &  &  &  &  &  & +  \\ \hline 
		\scriptsize{\textbf{PeerFinder} (\citet{Du2018})} & + &  &  &  &  &  &  &   & + & + & +  \\ \hline 
		\scriptsize{(\citet{Dominguez2018})} & + &  &  &  &  &  & + &  &  &  &   \\ \hline 
		\scriptsize{(\citet{Mueller2019})} & + & + &  &  & + &  &  &  &  &  & + \\ \hline 
		\scriptsize{\textbf{Mastery Grids} (\citet{Barria-Pineda2019})} & + &  &  &  &  &  & + &  & + & + & +   \\ \hline 
		\scriptsize{\textbf{ComBub} (\citet{Jin2018})} & + &  &  &  &  &  & + &  &  &  & +   \\\hline 
		\scriptsize{\textbf{CourseQ} (\citet{ma2021courseq})} &  &  &  & + &  &  & + &  &  &  &  +  \\\hline 
		\scriptsize{\textbf{BrainHood} (\citet{tsiakas2020brainhood})} & + &  &  &  &  &  &  &  &  & + & +   \\\hline 
		\scriptsize{\textbf{Rankfomsets} (\citet{bansal2020recommending})} & + &  &  &  &  &  &  &  &  &  & +   \\\hline 
		\scriptsize{\textbf{IW Browser} (\citet{narayanan2008towards})} &  & + &  &  & + &  &  &  &  &  &    \\\hline 
		\scriptsize{\textbf{uRank} (\citet{sciascio2017supporting})} & + &  &  &  &  &  & + &  &  &  &    \\\hline 
		\scriptsize{(\citet{alshammari2019mining})} & + &  &  &  &  &  & + &  &  &  &    \\\hline

	\end{tabular}
\end{table}

\subsubsection{What: Data Abstraction} \label{sec:survey-what}

\begin{figure}[h]
	\centering
	
	\begin{subfigure}[b]{0.48\textwidth}
		\includegraphics[width=\textwidth]{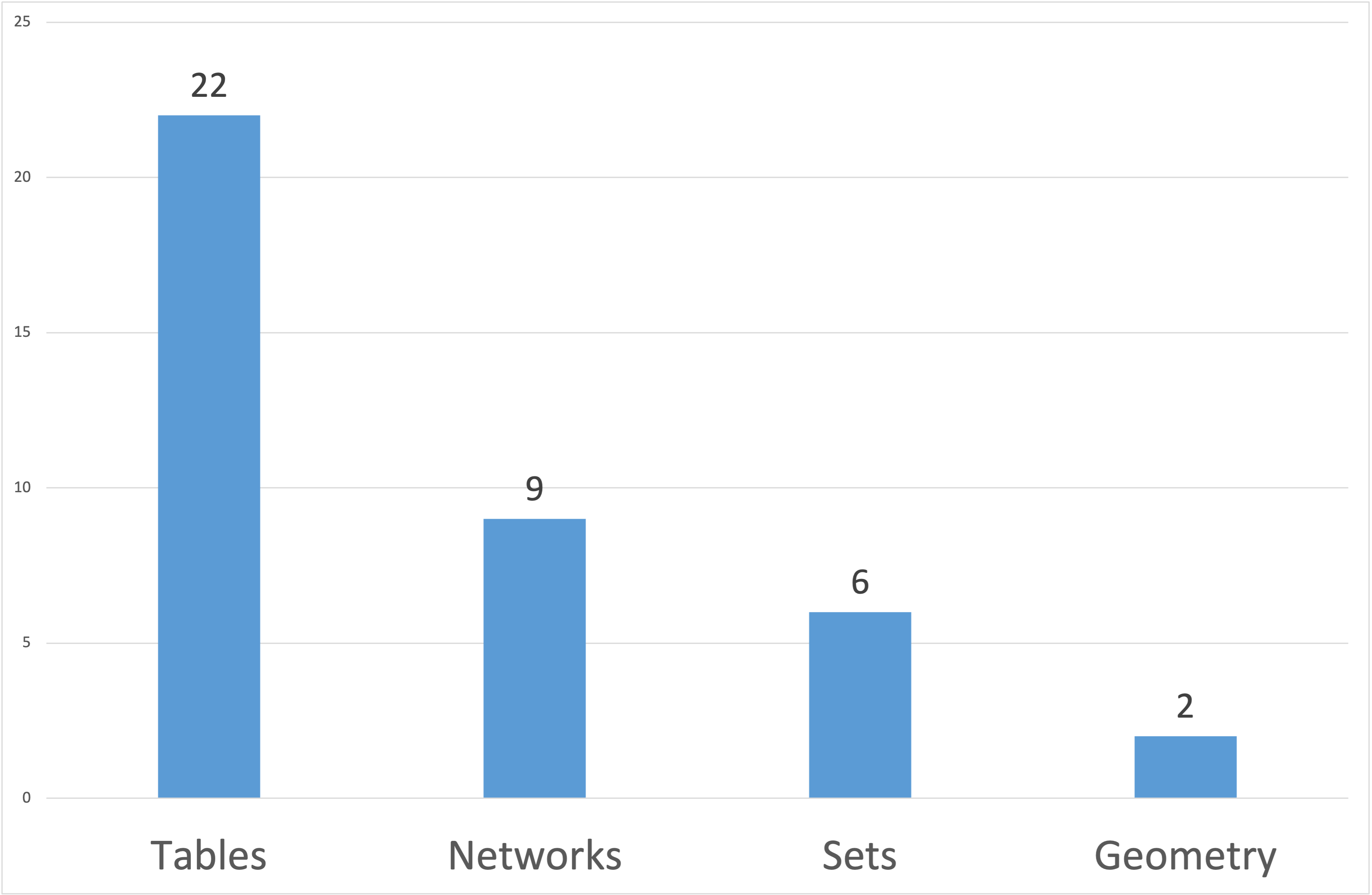}
		\caption{Data abstraction}
		\label{fig:survey-what}
	\end{subfigure}
	\hfill
	\begin{subfigure}[b]{0.48\textwidth}
		\includegraphics[width=\textwidth]{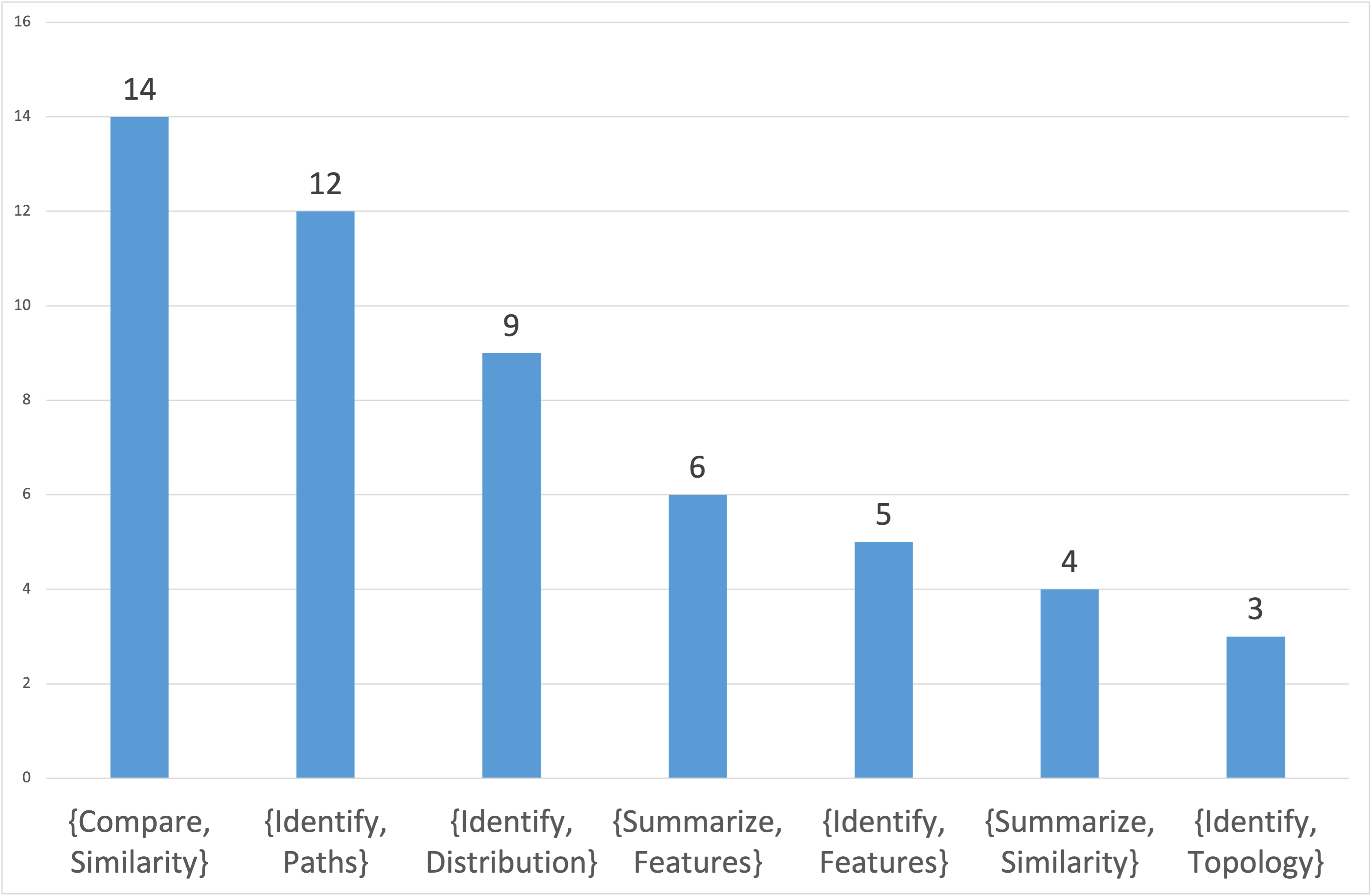}
		\caption{Task abstraction}
		\label{fig:survey-why}
	\end{subfigure}
	\caption{Explanation format (What / Why)}
	\label{fig2}
\end{figure}

The data used by the surveyed tools to generate visual explanations are categorized into four different dataset types, namely tables, networks, sets, and geometry (see Table \ref{table:survey-what-why} and  Figure \ref{fig:survey-what}). Tables are used in 22 tools to store different items/attributes, such as song attributes and user preferences for these attributes \cite{Millecamp2019}, artists and moods \cite{Andjelkovic2019}, publications, topics, co-authorships, and interests \cite{Tsai2019}, movies \cite{Symeonidis2009}, Facebook profile attributes and personality traits \cite{Jin2016}. Networks as a dataset type are mainly used in the tools providing recommendations based on social data. Examples of nodes/links in these networks include authors and co-authorships \cite{Tsai2019}, tags, recommender agents, and users \cite{Verbert2013}, music, Wikipedia items, Facebook friends, and Twitter experts \cite{Bostandjiev2012}, movies \cite{Vlachos2012}. Sets are used in many tools providing content-based explanation based on e.g., topics \cite{Tsai2019}, keywords \cite{Tsai2017,Tsai2019,Zhao2011}, feature-opinion word pairs \cite{Zhang2014}. Only two tools use the geometry dataset type. \citet{Tsai2019} use it to plot cities of affiliations on a world map, inspired by location-based explanation, and \citet{ma2021courseq} use it to plot courses representation on a two-dimensional layout to show the course relevance to the student interest.


\subsubsection{Why: Task Abstraction} \label{sec:survey-why}

As mentioned in Section \ref{subsubsec:vis-why}, a task abstraction is expressed as an \{action, target\} pair. 
12 out of the 33 surveyed tools are focusing on the \{identify, paths\} task (see Table \ref{table:survey-what-why} and Figure \ref{fig:survey-why}). These are mostly the tools that use networks as dataset type. For example, `Relevance Tuner+' uses a path graph to visualize different possible connections between two conference attendees based on co-authorship information \cite{Tsai2019}. A layer-based interface connected via outgoing links is used by `TasteWeights' and `LinkedVis' (three layers) as well as `SmallWorld' (five layers) to visualize connections between the user profile and the recommended items \cite{Bostandjiev2012,Bostandjiev2013,Gretarsson2010}. `PeerChooser' incorporates a force-directed graph layout technique to visualize connections between an active user and different user-communities/peer-groups \cite{ODonovan2008}.  Additionally, three of these tools focus on the \{identify, topology\} task to find clusters in the network based on user preferences \cite{Gretarsson2010,ODonovan2008,Verbert2013}. To provide explanations based on similarity of items or attributes, another common task in the surveyed tools is \{compare, similarity\} between liked and recommended music artists \cite{Kouki2019}, active user and recommended users as well as liked and recommended items \cite{Park2017}, users' song preferences and recommended songs \cite{Millecamp2019}, searched keyword and feedback \cite{Kangasraeaesioe2015} or \{summarize, similarity\} between all artists \cite{Kouki2019}, artist moods \cite{Andjelkovic2019}, research papers \cite{Parra2014}, topics discussed in content-centric Web sites \cite{Zhao2011}. The \{identify, features\} task is addressed by five tools, and the \{summarize, features\} task is addressed by six tools. For example, in \cite{Zhang2014}, the authors visualize the feature-opinion pairs of the recommended items as a tag cloud enabling users to identify the sentiment related to a specific pair of interests or get an overview of sentiments related to all the pairs. Similarly, in \cite{Bakalov2013}, the authors focus on both of these tasks to identify clusters and summarize interest degrees by visualizing items of interest categorized by different ontology classes on IntrospectiveViews which is divided into zones and slices. Nine tools focus on the \{identify, distribution\} task. `Tagsplanation' \cite{Vig2009}, for instance, shows the distribution of tags related to a specific movie in a bar chart. `System U' \cite{Badenes2014} presents the distribution of the personality traits of a user via a sunburst chart.

\subsubsection{How: Idioms} \label{sec:survey-how}
This section discusses the visualization idioms used in the surveyed tools. Different charts, such as node-link diagrams, tag clouds, and bar charts are used to encode the explanatory visualizations and various interaction techniques are provided to enable users to interact with and control the visualizations.

\begin{table}[h!] 
	\caption{Summary of the survey results - explanation format (How)}
	\label{table:survey-how}
	\setlength\tabcolsep{2.3pt}
	\begin{tabular}{|l|p{3.0mm}|p{3.0mm}|p{3.0mm}|p{3.0mm}|p{3.0mm}|p{3.0mm}|p{3.0mm}|p{3.0mm}|p{3.0mm}|p{3.0mm}|p{3.0mm}|p{3.0mm}|p{3.0mm}|p{3.0mm}|p{3.0mm}|p{3.0mm}|}
		\hline
		& 
		\multicolumn{16}{c|}{\scriptsize{\textbf{How}}} \\ \cline{2-17}
		
		& 
		\multicolumn{10}{c|}{\scriptsize{\textbf{Encode}}} & 
		\multicolumn{6}{c|}{\scriptsize{\textbf{Interact}}} \\ \cline{2-17}
		
		\multicolumn{1}{|c|}{\textbf{Tools}} & 
		\multicolumn{1}{c|}{\rotatebox{270}{\tiny{Node-Link Diagram}}} & 
		\multicolumn{1}{c|}{\rotatebox{270}{\tiny{Tag Cloud}}} & 
		\multicolumn{1}{c|}{\rotatebox{270}{\tiny{Bar Chart}}} & 
		\multicolumn{1}{c|}{\rotatebox{270}{\tiny{Venn Diagram}}} & 
		\multicolumn{1}{c|}{\rotatebox{270}{\tiny{Radar Chart}}} & 
		\multicolumn{1}{c|}{\rotatebox{270}{\tiny{World Map}}} & 
		\multicolumn{1}{c|}{\rotatebox{270}{\tiny{Heatmap}}} & 
		\multicolumn{1}{c|}{\rotatebox{270}{\tiny{Treemap}}} & 
		\multicolumn{1}{c|}{\rotatebox{270}{\tiny{Piechart}}} & 
		\multicolumn{1}{c|}{\rotatebox{270}{\tiny{Scatterplot}}} &  
		
		\multicolumn{1}{c|}{\rotatebox{270}{\tiny{Manipulate: Select}}} & 
		\multicolumn{1}{c|}{\rotatebox{270}{\tiny{Reduce: Filter}}} & 
		\multicolumn{1}{c|}{\rotatebox{270}{\tiny{Facet: Juxtapose}}} & 
		\multicolumn{1}{c|}{\rotatebox{270}{\tiny{Manipulate: Change}}} & 
		\multicolumn{1}{c|}{\rotatebox{270}{\tiny{Manipulate: Navigate }}} & 
		\multicolumn{1}{c|}{\rotatebox{270}{\tiny{No Interaction}}} \\ \hline
		
		\tiny{\textbf{Relevance Tuner+} (\citet{Tsai2019})} &  + & + & + & + & + & + &  &  &  &  + &  &  & + & + & & \\ \hline 
		\tiny{\textbf{HyPER} (\citet{Kouki2019})} & + & + &  & + &  &  &  &  &  &  + &  &  &  &  & & \\ \hline 
		\tiny{\textbf{Tagsplanation} (\citet{Vig2009})} &  &  & + &  &  &  &  &  &  &   &  &  &  &  & + & \\ \hline  
		\tiny{\textbf{PeerChooser} (\citet{ODonovan2008})} & + &  &  &  &  &  &  &  &  &  + & + & + & + &  &  & \\ \hline 
		\tiny{\textbf{TalkExplorer} (\citet{Verbert2013})} & + &  &  &  &  &  &  &  &  &  + & + & + &  &  &  & \\ \hline 
		\tiny{\textbf{SetFusion} (\citet{Parra2014})} &  &  &  & + &  &  &  &  &  &   & + & + &  &  &  & \\ \hline 
		\tiny{\textbf{TasteWeights} (\citet{Bostandjiev2012})} & + &  &  &  &  &  &  &  &  &  + & + &  &  &  & &  \\ \hline  
		\tiny{\textbf{Moodplay} (\citet{Andjelkovic2019})} &  & + &  &  &  &  &  &  &  &  + & + & + &  & + &  & \\ \hline 
		\tiny{\textbf{SmallWorld} (\citet{Gretarsson2010})} & + &  &  &  &  &  &  &  &  &  + &  &  &  & + &  &  \\ \hline 
		\tiny{\textbf{LinkedVis} (\citet{Bostandjiev2013})} & + &  &  &  &  &  &  &  &  &  + & + &  &  &  &  & \\ \hline 
		\tiny{(\citet{Schaffer2015})} & + &  &  &  &  &  &  &  &  &  + & + &  &  &  &  & \\ \hline  
		\tiny{(\citet{Vlachos2012})} & + &  &  &  &  &  &  &  &  &  + & + & + &  &  &  & \\ \hline 
		\tiny{\textbf{PARIS-ad} (\citet{Jin2016})} & + &  &  &  &  &  &  &  &  &  + & + & + &  &  & &  \\ \hline 
		\tiny{(\citet{Bakalov2013})} &  &  &  &  &  &  & + &  &  &  &  + & + &  &  &  & \\ \hline 
		\tiny{(\citet{Kangasraeaesioe2015})} &  & + &  &  &  &  &  &  &  &  + &  &  & + &  & & \\ \hline 
		\tiny{\textbf{System U} (\citet{Badenes2014})} &  &  &  &  &  &  &  & + &  &  + &  & + &  &  & & \\ \hline 
		\tiny{\textbf{UniWalk} (\citet{Park2017})} &  &  & + &  &  &  &  &  &  &  + &  & + &  &  & & \\ \hline 
		\tiny{(\citet{Millecamp2019})} &  &  & + &  &  &  &  &  & + &  + & + & + & + &  & & \\ \hline 
		\tiny{\textbf{RelExplorer} (\citet{Tsai2017})} &  + & + &  &  &  &  &  &  &  &  & + &  &  &  & & \\ \hline 
		\tiny{\textbf{Pharos} (\citet{Zhao2011})} &  & + &  &  &  &  &  &  &  &  + &  & + &  &  & &  \\ \hline 
		\tiny{(\citet{Zhang2014})} &  & + &  &  &  &  &  &  &  &   &  &  &  &  & + & \\ \hline 
		\tiny{\textbf{LIBRA} (\citet{Bilgic2005})} &  &  & + &  &  &  &  &  &  &   &  &  &  &  & + & \\ \hline 
		\tiny{\textbf{PeerFinder} (\citet{Du2018})} &  &  & + &  &  &  & + & + &  &  + & + & + &  &  & &  \\ \hline 
		\tiny{(\citet{Dominguez2018})} &  &  & + &  &  &  &  &  &  &   &  &  &  &  & + & \\ \hline 
		\tiny{(\citet{Mueller2019})} & + &  & + &  &  &  &  &  &  &  + & + & + &  & + &  & \\ \hline 
		\tiny{\textbf{Mastery Grids} (\citet{Barria-Pineda2019})} &  &  & + &  &  &  & + &  &  &  + &  & + &  &  & &  \\ \hline 
		\tiny{\textbf{ComBub} (\citet{Jin2018})} &  &  &  &  &  &  &  &  & + & + & + & + &  &  &  & \\\hline	
		\tiny{\textbf{CourseQ} (\citet{ma2021courseq})} &  &  &  &  &  &  &  &  &  & + & + & + &  &  & + &  \\\hline
		\tiny{\textbf{BrainHood} (\citet{tsiakas2020brainhood})} &  &  & + &  &  &  &  &  & + &  & + &  & + &  &  &  \\\hline
		\tiny{\textbf{Rankfomsets} (\citet{bansal2020recommending})} &  &  & + &  &  &  &  &  &  &  & + &  & + &  &  &  \\\hline
		\tiny{\textbf{IW Browser} (\citet{narayanan2008towards})} & + &  &  &  &  &  &  &  &  &  &  &  &  & & + &  \\\hline
		\tiny{\textbf{uRank} (\citet{sciascio2017supporting})} &  &  & + &  &  &  &  &  &  &  & + & + & + &  & + &  \\\hline
		\tiny{(\citet{alshammari2019mining})} &  &  & + &  &  &  &  &  &  &  &  &  &  &  & & + \\ \hline

	\end{tabular}
\end{table}

\begin{figure}[ht!]
	\centering
	
	\begin{subfigure}[b]{0.48\textwidth}
		\includegraphics[width=\textwidth]{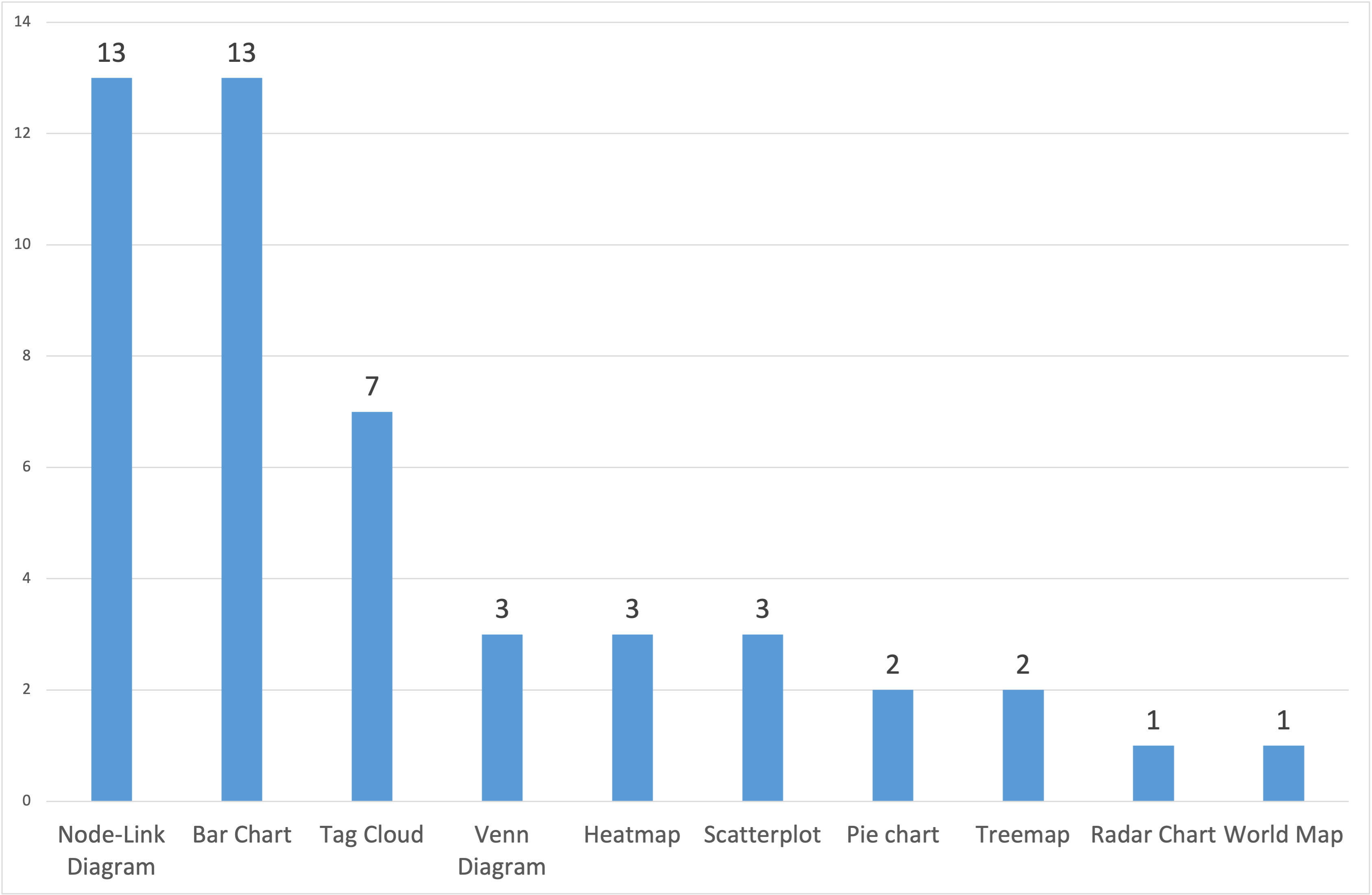}
		\caption{Chart types}
		\label{fig:survey-how-encode}
	\end{subfigure}
	\hfill
	\begin{subfigure}[b]{0.48\textwidth}
		\includegraphics[width=\textwidth]{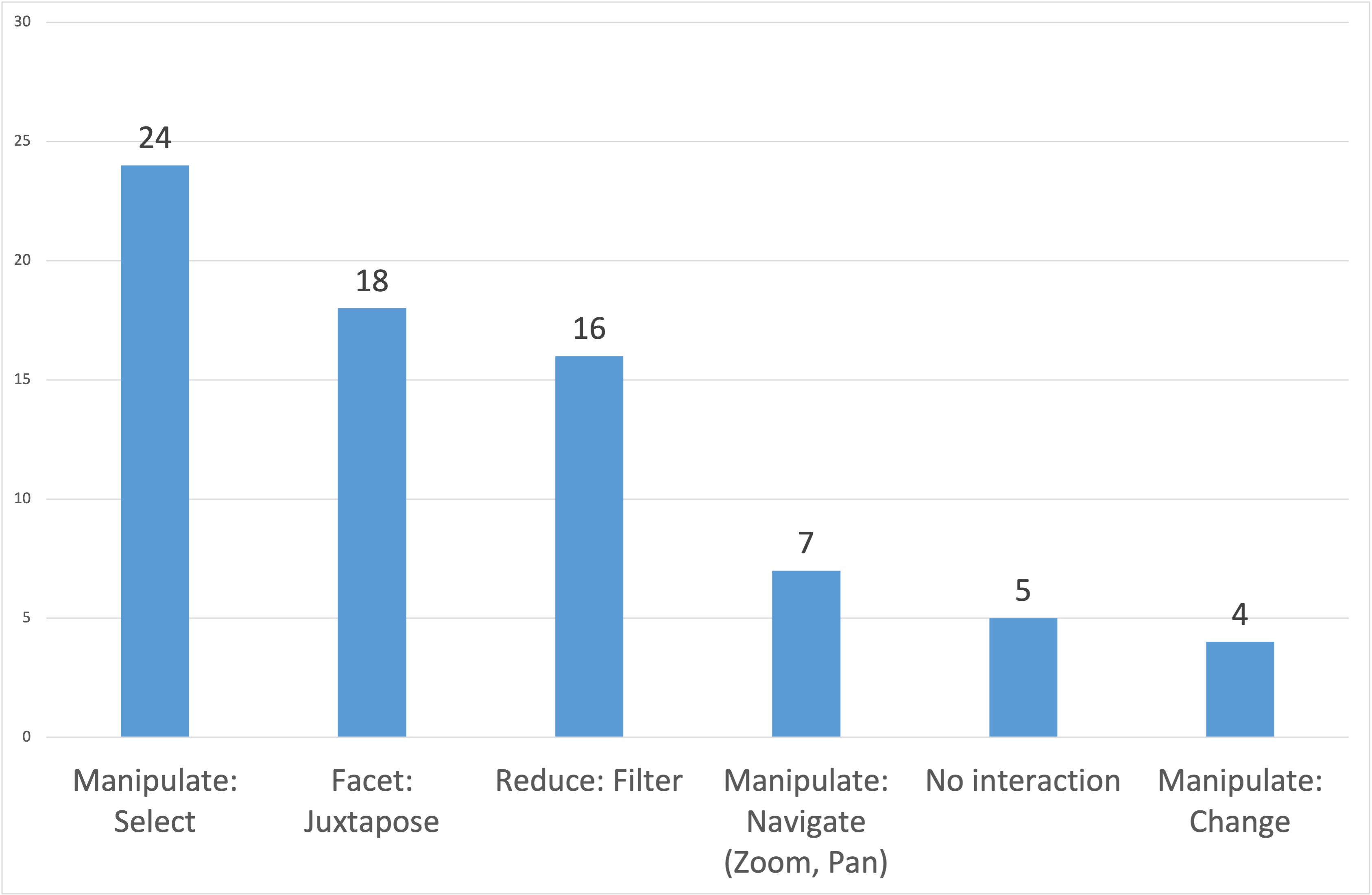}
		\caption{Interactivity types}
		\label{fig:survey-how-interact}
	\end{subfigure}
	\caption{Explanation format (How)}
	\label{fig3}
\end{figure}
\begin{figure}[htb!]
	\centering
	\begin{subfigure}[b]{0.48\textwidth}
		\includegraphics[width=\textwidth]{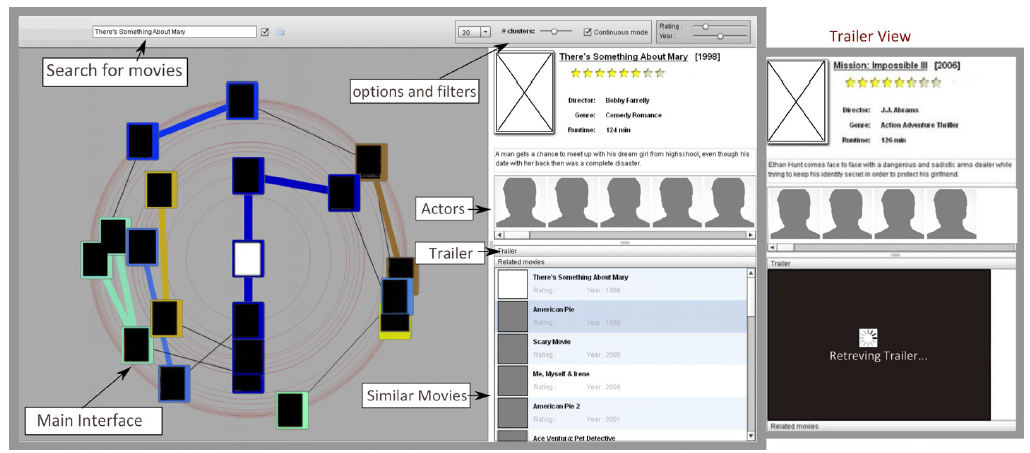}
		\caption{\citet{Vlachos2012} visualize the connection between movies based on their similarity.}
		\label{fig:vlachos12}
	\end{subfigure}
	\hfill
	\begin{subfigure}[b]{0.48\textwidth}
		\includegraphics[width=\textwidth]{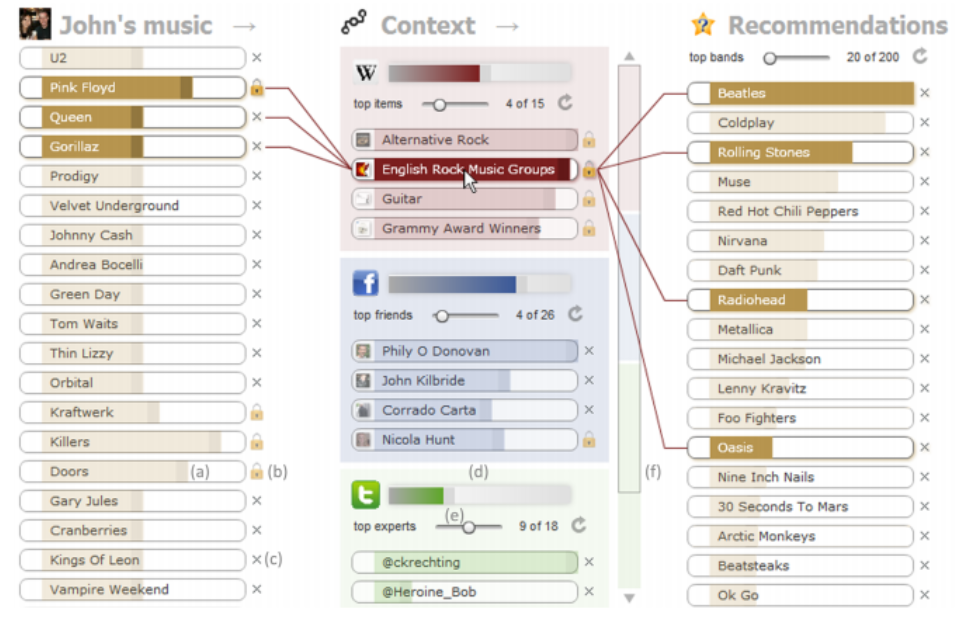}
		\caption{TasteWeights \cite{Bostandjiev2012} visualizes connections between the user profile, context, social commection, and the recommended music to explain the recommendation.} 
		\label{fig:Bostandjiev12_TasteWeights}
	\end{subfigure}
	\hfill
	\begin{subfigure}[b]{0.48\textwidth}
		\includegraphics[width=\textwidth]{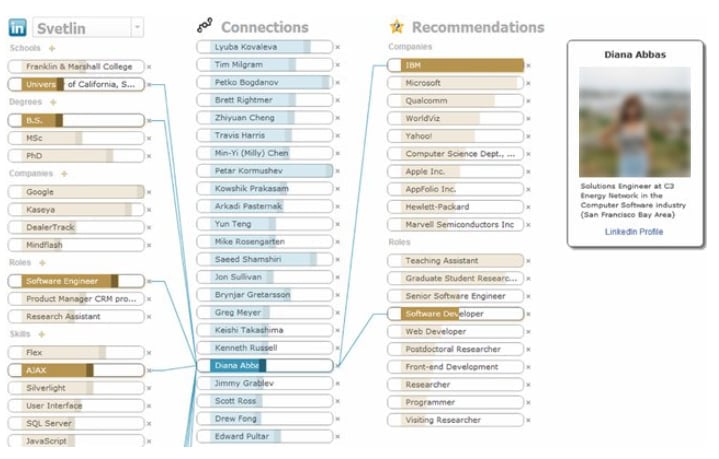}
		\caption{LinkedVis \cite{Bostandjiev2013} explains the recommendation by highlighting the relationships between user profile attributes (LinkedIn), social connections, and the recommended items (company and role). }
		\label{fig:Bostandjiev13_linkedVis}
	\end{subfigure}
	\hfill
	\begin{subfigure}[b]{0.48\textwidth}
		\includegraphics[width=\textwidth]{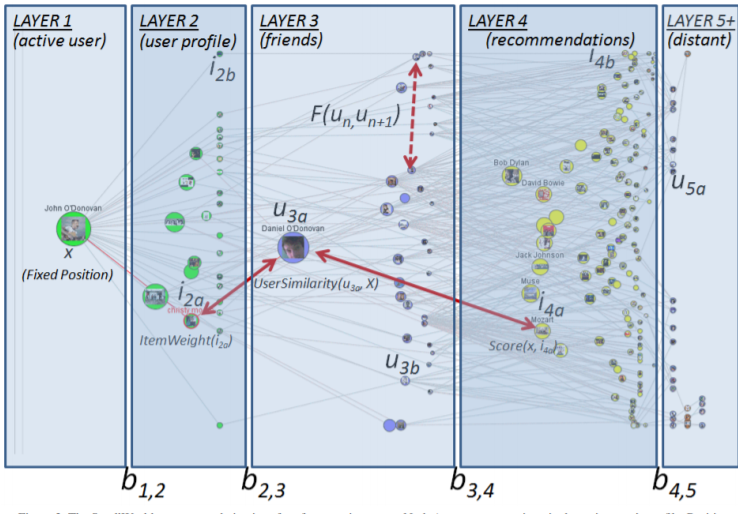}
		\caption{SmallWorld \cite{Gretarsson2010} organizes related items into layers of nodes to explain the recommendation through connections.}
		\label{fig:Gretarsson10_SmallWorlds}
	\end{subfigure}
	\hfill
	\begin{subfigure}[b]{0.48\textwidth}
		\includegraphics[height=0.26\textheight, width=\textwidth]{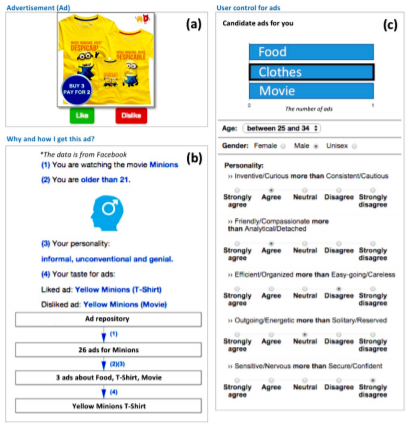}
		\caption{PARIS-ad \cite{Jin2016} uses a basic node-link diagram to illustrate the recommendation process based on the user profile.}
		\label{fig:Jin16_ParisAd}
	\end{subfigure}
	\caption{Tools using a node-link diagram}
	\label{fig4}
\end{figure}

\textbf{Encode}
To encode the explanatory visualizations, different charts are used by the tools that we have surveyed (see Table \ref{table:survey-how} and Figure \ref{fig:survey-how-encode}). The most commonly used visualization idioms are the node-link diagram and the bar chart which are used each by 13 out of the 27 surveyed tools. The node-link diagram idiom is used to indicate the relationships between items. For instance, the tool in \cite{Vlachos2012} represents movies as nodes and visualizes the connection between them based on their similarity (Figure \ref{fig:vlachos12}). The tool in \cite{Schaffer2015}, `TasteWeights' \cite{Bostandjiev2012}, `LinkedVis' \cite{Bostandjiev2013}, and `SmallWorld' \cite{Gretarsson2010} aggregate related items representing nodes in the form of parallel layers and visually explain the resulted recommendations via connections (Figure \ref{fig:Bostandjiev12_TasteWeights}, \ref{fig:Bostandjiev13_linkedVis}, \ref{fig:Gretarsson10_SmallWorlds}). `PARIS-ad' \cite{Jin2016} uses a simple node-link diagram consisting of four nodes to explain the recommendation process based on the user profile (Figure \ref{fig:Jin16_ParisAd}). Further, we observed that node-link diagrams are favorable idioms in tools that use the social explanation method (eight tools). To summarize, the node-link diagram is an effective explanatory visualization for network-based datasets to enable efficient path identification and topology exploration.


\begin{figure}[hbt!]
	\centering
	
	\begin{subfigure}[b]{0.38\textwidth}
		\includegraphics[width=\textwidth]{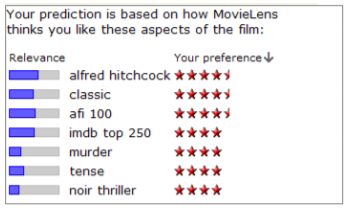}
		\caption{Tagsplanation \cite{Vig2009} uses bar charts to show tag relevance in movies and user preferences.}
		\label{fig:Vig9_Tagsplanations}
	\end{subfigure}
	\hfill
	\begin{subfigure}[b]{0.58\textwidth}
		\includegraphics[height=0.2\textheight, width=\textwidth]{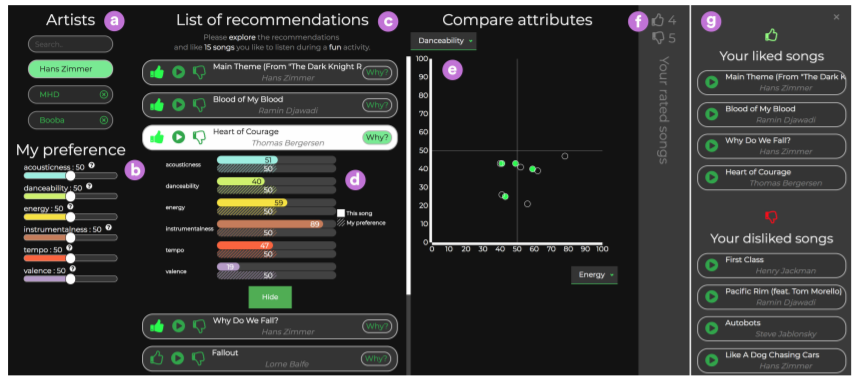}
		\caption{\citet{Millecamp2019} use a grouped bar chart to compare recommended song attributes with user preferences.}
		\label{fig:millecamp19}
	\end{subfigure}
	\hfill
	\begin{subfigure}[b]{0.8\textwidth}
		\includegraphics[height=0.15\textheight,width=\textwidth]{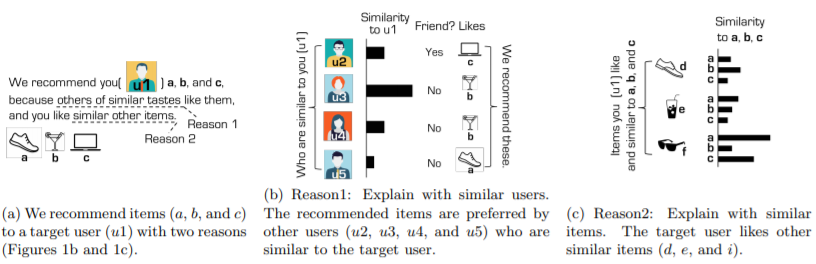}
		\caption{UniWalk \cite{Park2017} uses bar charts to explain recommended items based on similar users and similar items.}
		\label{fig:park17_UniWalk}
	\end{subfigure}
	\caption{Tools using a bar chart}
	\label{fig5}
\end{figure}

The bar chart idiom is mainly used to compare similarity between user preferences and recommendations. For instance, to justify the recommended movie based on community tags, `Tagsplanation' \cite{Vig2009} uses bar charts to visualize each tag's relevance in the movie and the user's preference (Figure \ref{fig:Vig9_Tagsplanations}). The tool in \cite{Millecamp2019} uses a grouped bar chart to show a comparison between the recommended song's attributes and user's preferences (Figure \ref{fig:millecamp19}). Similarly, `UniWalk' \cite{Park2017} uses bar charts to provide an explanation of recommended items based on similar users who liked those items and similar items liked by the user (Figure \ref{fig:park17_UniWalk}). Overall, bar chart is one of the most popular techniques to provide comparisons and similarities in table-based datasets. 


\begin{figure}[hbt!]
	\centering
	
	\begin{subfigure}[b]{0.48\textwidth}
		\includegraphics[width=\textwidth]{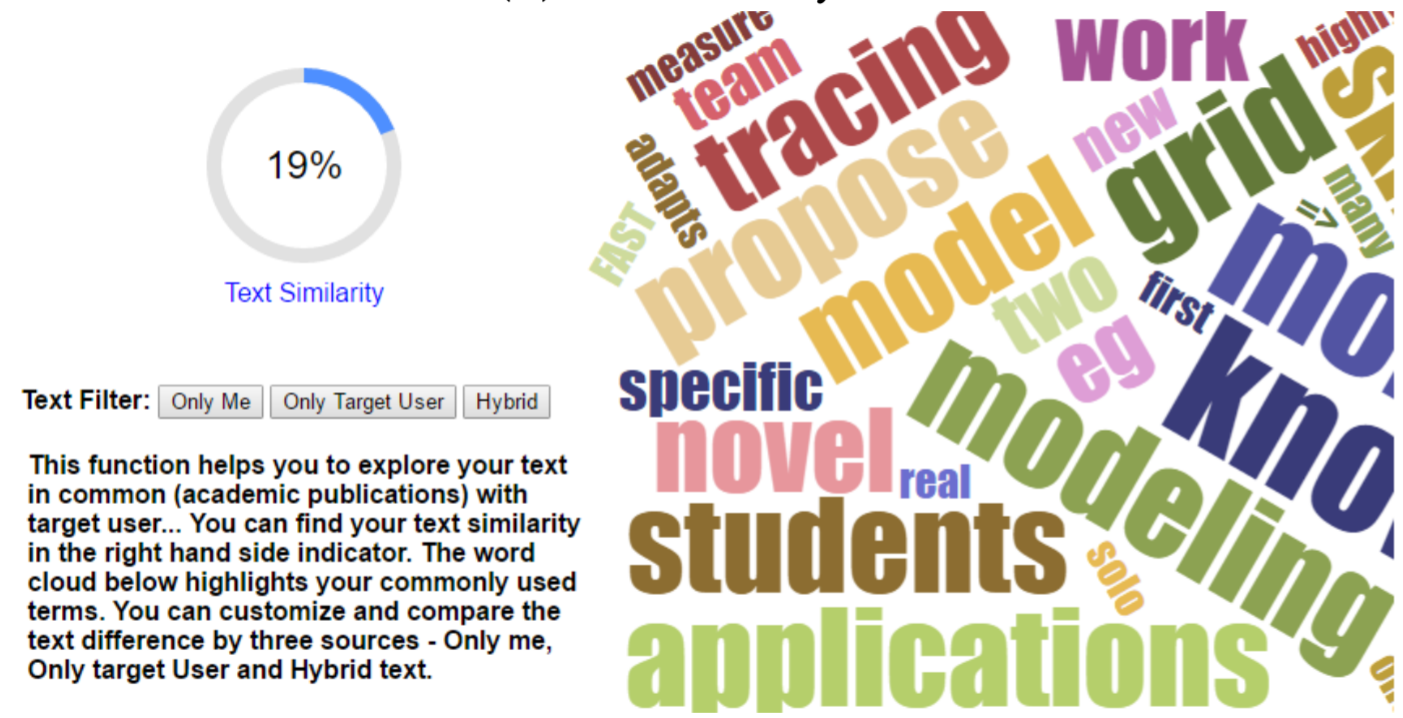}
		\caption{RelExplorer \cite{Tsai2017} uses a tag cloud to help conference attendees assess content similarity between their publications and those of target scholars.}
		\label{fig:Tsai17_RelExplorer}
	\end{subfigure}
	\hfill
	\begin{subfigure}[b]{0.48\textwidth}
		\includegraphics[width=\textwidth]{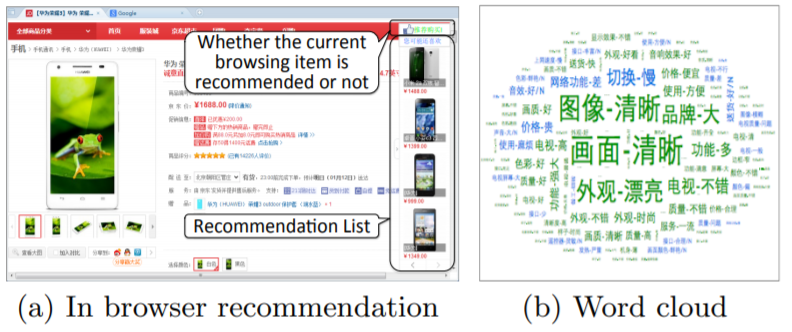}
		\caption{ \citet{Zhang2014} explain recommendations using feature-opinion word pairs, distinguished by positive (green) and negative (blue) sentiments.}
		\label{fig:Zhang14_wordcloud}
	\end{subfigure}
	\hfill
	\begin{subfigure}[b]{0.48\textwidth}
		\includegraphics[width=\textwidth]{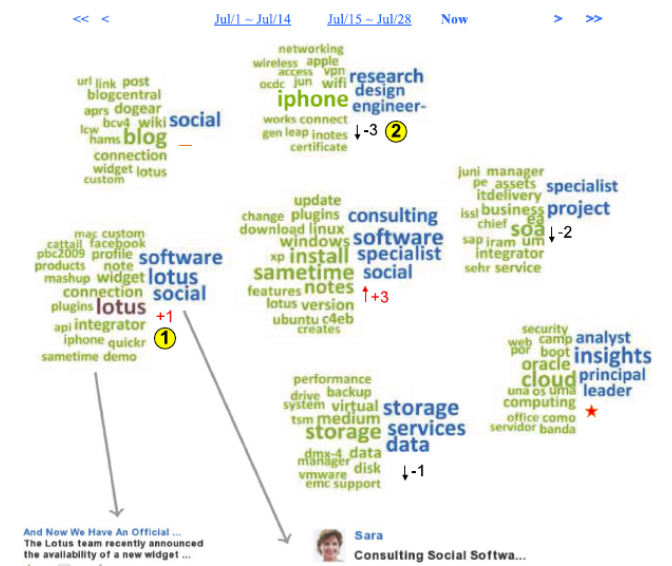}
		\caption{Pharos \cite{Zhao2011} uses tag clouds to summarize users' social behavior, with topics visualized in green and people in blue.}
		\label{fig:Zhao11_Pharos}
	\end{subfigure}
	\caption{Tools using a tag cloud}
	\label{fig6}
\end{figure}

Tag cloud is the third most used visualization idiom, which is used by seven surveyed tools to represent textual data. This idiom is mostly used in tools that use the content-based explanation method. `RelExplorer' \cite{Tsai2017} uses a tag cloud to assist co-attendees of an academic conference in understanding the level of content similarity between their publications and those of target scholars (Figure \ref{fig:Tsai17_RelExplorer}). Similarly, the tool in \cite{Zhang2014} provides the explanation of recommendations in the form of feature-opinion word pairs, which are further distinguished based on positive (green) and negative (blue) sentiments (Figure \ref{fig:Zhang14_wordcloud}). `Pharos' \cite{Zhao2011} uses a set of tag clouds, representing communities focusing on a similar topic, to summarize users' social behavior in content-centric Web sites, where each tag cloud visualizes the topics in green and people in blue (Figure \ref{fig:Zhao11_Pharos}). In general, tag clouds are suitable for summarizing or comparing feature similarity in set-based datasets.

\begin{figure}[hbt!]
	\centering
	
	\begin{subfigure}[b]{0.29\textwidth}
		\includegraphics[width=\textwidth]{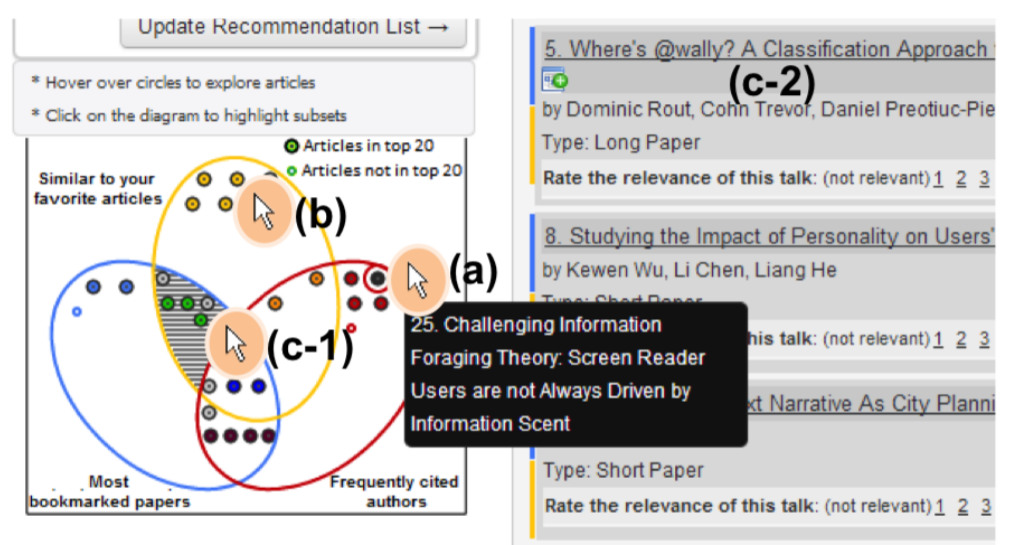}
		\caption{SetFusion \cite{Parra2014} visualizes recommended talks related to conference publications common among three sets: user's favorite, most bookmarked, and frequently cited authors.}
		\label{fig:Parra14_VennD}
	\end{subfigure}
	\hfill
	\begin{subfigure}[b]{0.29\textwidth}
		\includegraphics[width=\textwidth]{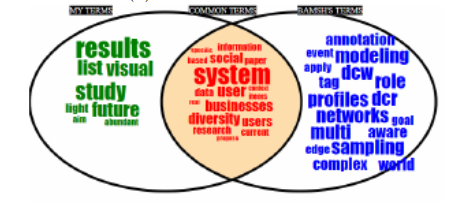}
		\caption{Relevance Tuner+ \cite{Tsai2019} uses a Venn diagram and tag cloud to illustrate the similarity between a user's publications and those of a conference attendee.}
		\label{fig:Tsai19_VennD}
	\end{subfigure}
	\hfill
	\begin{subfigure}[b]{0.29\textwidth}
		\includegraphics[width=\textwidth]{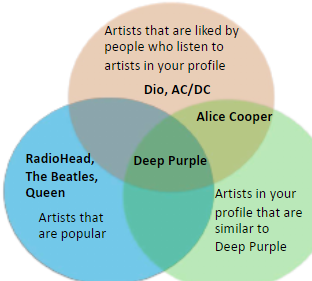}
		\caption{HyPER \cite{Kouki2019} recommends intersecting music artists between user profile, popular artists, and those liked by users with similar tastes using a Venn diagram, and employs a tag cloud in each circle displaying set-specific keywords.}
		\label{fig:kouki19_VennD}
	\end{subfigure}
	
	\caption{Tools using a Venn diagram}
	\label{fig7}
\end{figure}

The remaining seven visualization idioms, namely Venn diagram, heatmap, treemap, scatterplot, radar chart, pie chart, and world map are used less frequently in the tools that we reviewed. Three tools use a Venn diagram to show possible intersections between their datasets. 'SetFusion' \cite{Parra2014} uses it to visualize recommended talks related to the publications in a conference which are common between three different sets containing publications similar to the user's favorite, most bookmarked, and from frequently cited authors (Figure \ref{fig:Parra14_VennD}). Similarly, `Relevance Tuner+' \cite{Tsai2019} uses a Venn diagram together with a tag cloud to explain the similarity between the publications of a user and attendee of a conference (Figure \ref{fig:Tsai19_VennD}). `HyPER' \cite{Kouki2019} also uses a Venn diagram to recommend intersecting music artists between sets of artists from a user profile, popular artists, and artists liked by people listening to similar artists as the user. Additionally, `HyPER' uses a tag cloud in each circle of the Venn diagram to visualize keywords from a particular set (Figure \ref{fig:kouki19_VennD}). To sum up, Venn diagrams alone are good for comparing items and features, but together with tag clouds, they can provide effective visualizations to explain content-based datasets.

\begin{figure}[hbt!]
	\centering
	\begin{subfigure}[b]{0.48\textwidth}
		\includegraphics[width=\textwidth]{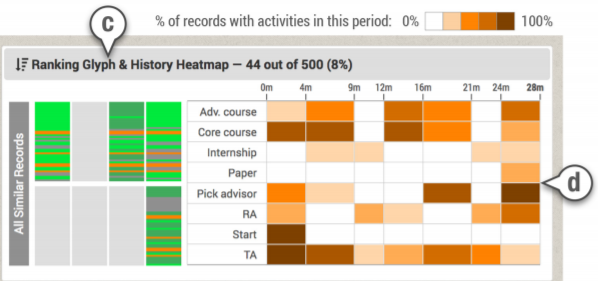}
		\caption{PeerFinder \cite{Du2018} summarizes temporal activities and events among similar users using an orange color gradient.}
		\label{fig:Du18_heatmap}
	\end{subfigure}
	\hfill
	\begin{subfigure}[b]{0.48\textwidth}
		\includegraphics[width=\textwidth]{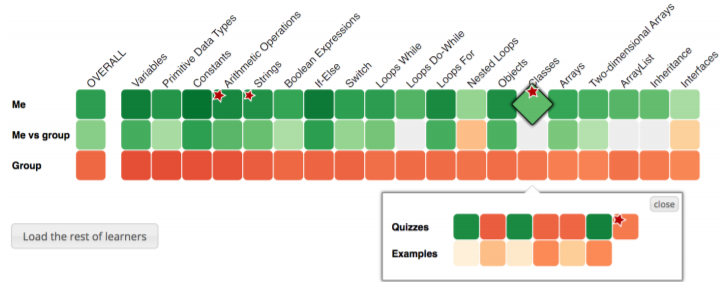}
		\caption{Mastery Grids \cite{Barria-Pineda2019} uses a heatmap to estimate student mastery intensity of domain concepts.}
		\label{fig:pineda19_heatmap}
	\end{subfigure}
	\caption{Tools using a heatmap}
	\label{fig8}
\end{figure}

Heatmaps are used in three tools to identify and summarize interesting items. A radial heatmap called IntrospectiveViews is used in the tool proposed in \cite{Bakalov2013} to visualize items based on the degree of interest and the type of item recommended. Most interesting items are displayed in the central hot zone while the least interesting items are presented in the cold zone on the edge of the radial heatmap. A history heatmap is provided in 'PeerFinder' \cite{Du2018} to summarize the temporal activities and events among similar users using an orange color gradient (Figure \ref{fig:Du18_heatmap}). In 'Mastery Grids' \cite{Barria-Pineda2019}, a heatmap is used to estimate the intensity of the student mastery of domain concepts (Figure \ref{fig:pineda19_heatmap}).

\begin{figure}[hbt!]
	\centering

	\begin{subfigure}[b]{0.45\textwidth}
		\includegraphics[height=0.2\textheight, width=\textwidth]{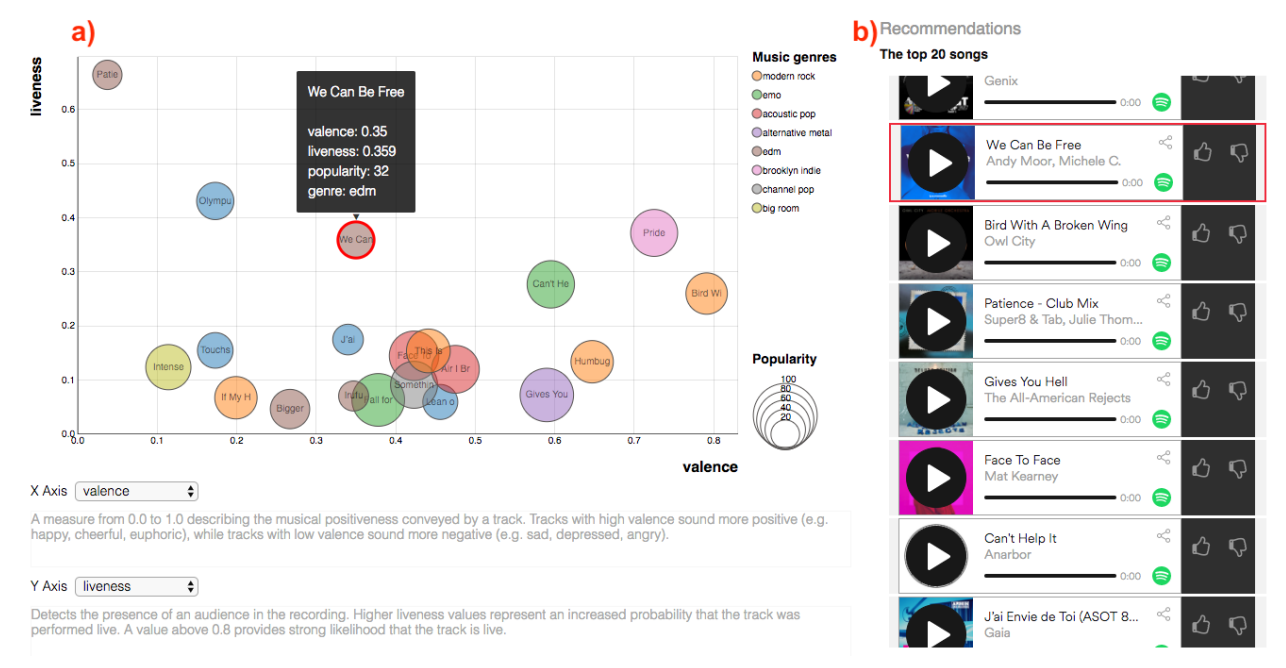}
		\caption{Combus \cite{Jin2018} uses a scatterplot to show recommended songs based on specified audio features, with bubbles representing popularity score and genre.} 
		\label{fig:Jin18_scatter}
	\end{subfigure}
	\hfill
	\begin{subfigure}[b]{0.45\textwidth}
		\includegraphics[height=0.2\textheight,width=\textwidth]{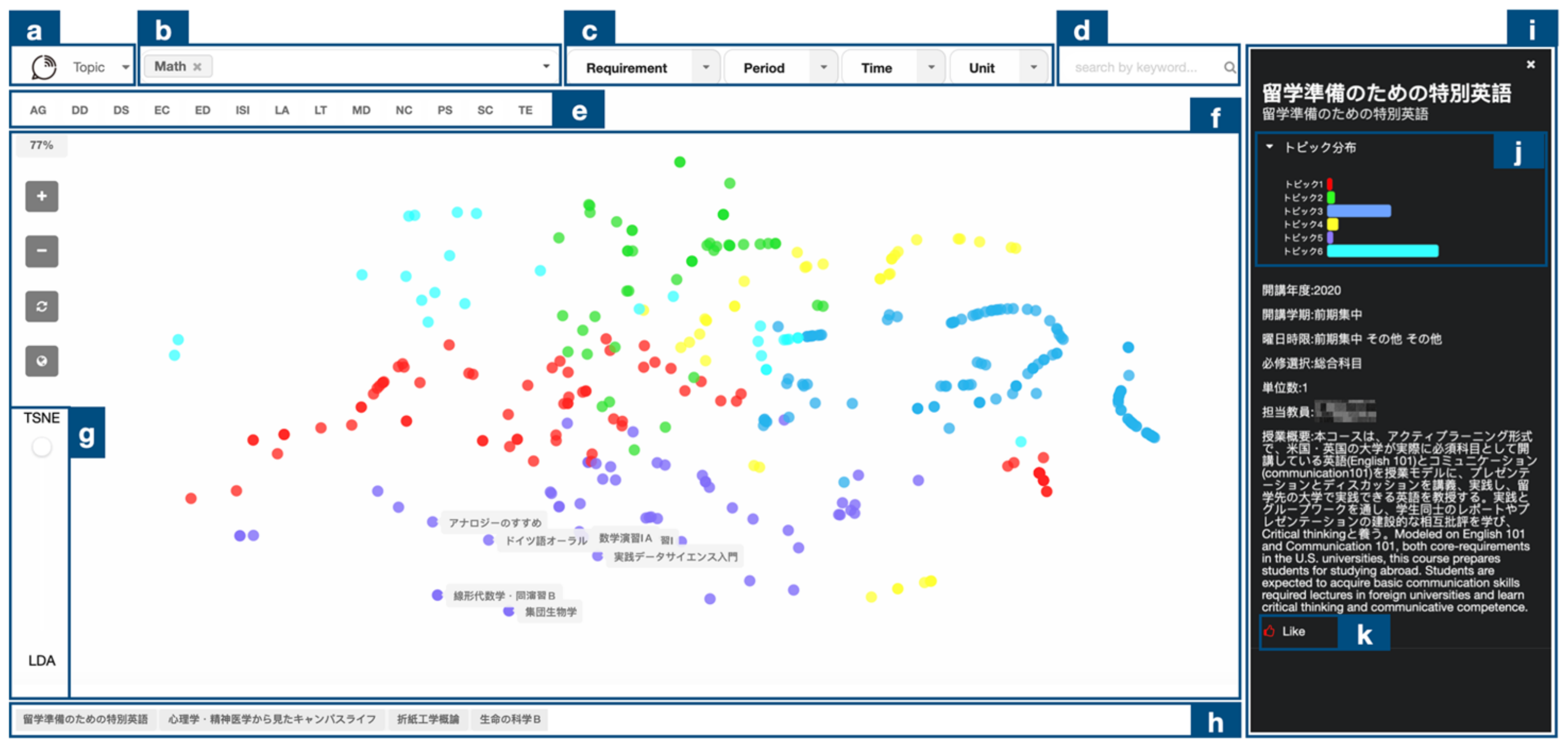}
		\caption{ CourseQ \cite{ma2021courseq} uses a dynamic scatterplot for students to explore recommended courses based on their interests, with coloring and highlighting features.}
		\label{fig:Ma21_scatter}
	\end{subfigure}
	\caption{Tools using a scatterplot}
	\label{fig9}
\end{figure}

Scatterplots are also used in three tools as an intuitive way to show multidimensional data. To compare different attributes of all the recommended songs, \citet{Millecamp2019} use a scatterplot to display these songs in a two-dimensional plot where users can select the attributes they wish to see on the X and Y axes (Figure \ref{fig:millecamp19}). The green points represent the liked songs by the users, and the user’s attribute preference is shown by two lines forming a cross in the middle. 'ComBus' \cite{Jin2018} uses a scatterplot with two specified audio features to plot the recommended songs where they are considered as bubbles with different sizes and colors representing the popularity score and the genre, respectively (Figure \ref{fig:Jin18_scatter}). In 'CourseQ' \cite{ma2021courseq}, a dynamic scatterplot is used to help students explore and understand the recommended courses based on their interests using coloring and highlighting features (Figure \ref{fig:Ma21_scatter}).

\textbf{Interact}
The majority of the surveyed tools provide interactivity to help users interact with the explanatory visualizations (see Table \ref{table:survey-how} and Figure \ref{fig:survey-how-interact}). To manipulate views, 24 out of the 33 surveyed tools provide the ability to select an element to provide more details \cite{Kouki2019,Tsai2019,tsiakas2020brainhood,sciascio2017supporting}, highlight network paths \cite{Bostandjiev2012,Bostandjiev2013,Gretarsson2010,Schaffer2015,bansal2020recommending}. Four tools allow users to change the view by moving a selected keyword to bring similar items closer \cite{Kangasraeaesioe2015}, adding and removing network nodes to update recommendations and rearrange network \cite{ODonovan2008}. Seven tools implemented the zoom and pan functionality e.g., to view cities of affiliations on a world map \cite{Tsai2019} or to more closely inspect areas of interest on mood space visualization \cite{Andjelkovic2019}. To support the reduce interactions, 16 tools incorporated different filtering mechanisms to modify the visualization view using checkboxes \cite{Andjelkovic2019,Bakalov2013,Verbert2013}, sliders \cite{Bakalov2013,Bostandjiev2012,Bostandjiev2013,Millecamp2019,Parra2014,Schaffer2015}, buttons \cite{Bostandjiev2012,Bostandjiev2013,ODonovan2008,Schaffer2015,sciascio2017supporting}, or dropdown menu \cite{Bakalov2013,Millecamp2019,ma2021courseq}. 18 tools focused on facet interactions by juxtaposing and coordinating multiple views through selecting a topic to show recommended content and people \cite{Zhao2011}, a recommended user or item to show explanation \cite{Park2017}, or a region on a global view to get a detailed view \cite{Bakalov2013}.

\section{Discussion} \label{sec:discussion}
Our study aims to shed light on an under-explored area in the context of explainable recommendation, namely investigating the literature on explainable recommendation from the angle of explanatory visualization. Based on the results of our review, we derive insights and guidelines that might be constructive for designing visual explanations in recommender systems. Additionally, we have identified certain gaps that may lead to future lines of work in this field. The discussion section ends
with a summary of suggestions for effective visual explanation design (D) and
future research (R) that are referenced through the whole section, e.g.,
D1, R2.

\subsection{Explanation Aim}
Our analysis revealed that transparency (i.e., providing the reasoning behind the recommendation algorithm) is underexplored in the tools that we investigated, as compared to justification (i.e., providing an abstract description of how recommendations are generated). Opting for presenting more user-oriented justifications, rather than offering transparency by explaining the rationale of
the recommendation algorithm is common in the explainable recommendation literature \cite{balog2019, Gedikli2014, Vig2009, zhang2020explainable}. 
This is mainly due to the fact that in contemporary recommender systems, the underlying algorithm is often too complex or not intuitive to be described in a human-interpretable manner (e.g., deep learning models), or may involve details that the recommender system provider wishes to protect. In some situations where transparency is preferred (e.g., to help experienced users and developers understand, debug, and refine the recommendation algorithm), there is a need to adopt and adapt current research on leveraging visualizations for AI/machine learning explainability in the explainable recommendation domain (see Section \ref{sec:vis_exp}). 

Furthermore, we observed that most of the studies analyzed in our review investigated the effects of one single visual explanation on one or more explanation aims. However, evaluating multiple visual explanations in parallel and exploring their effects on different explanation aims is under-investigated. Given that the studies concern different visual explanations, the reported partly contradicting results mean that different visual explanations have different effects on different explanation aims. For instance, targeting justification or scrutability does not require the same kind or amount of visual explanation information compared to targeting transparency. Further, a complete and sound visual explanation may lead to increased objective transparency but not necessarily to more user-perceived transparency, satisfaction, or trust in the recommender system. This suggests that it is important to consider the intended explanation aim when designing a visual explanation, as also highlighted in previous research on explainable recommendation (e.g., \cite{Tintarev2012, Tintarev2015, Nunes2017}). Thus, an important direction for future work can be to conduct systematic evaluations aiming at answering the question ``what kind of visual explanation is more effective for what kind of explanation aim?'' \textbf{(R1)}.
\subsection{Explanation Scope}
The majority of the reviewed tools focus on explaining the output of the recommendation (i.e., justification) rather than explaining the process by providing insights into how the recommendation process works (i.e., transparency) or explaining the input by opening user models and allowing users to modify them (i.e., scrutability). Often, explaining the process (i.e., transparency) is not a straightforward task, given that the underlying algorithm might be too complex to be described in a human-interpretable manner. And, in many cases, just explaining the output (i.e., justification) might be enough. Ideally, a recommender system should provide all three options, i.e., explaining the output, the process, and the input, and then let users select the appropriate explanation based on their needs and preferences. 

Moreover, there is a need for more work that leverages interactive visualizations to help users scrutinize their models and see the effect on the recommendation output. Transparent, scrutable, and explainable user models would help users effectively and efficiently personalize and correct the recommendation. A systematic analysis of the effects of opening, scrutinizing, and explaining user models using visualization represents an important direction for future research \textbf{(R2)}.

\subsection{Explanation Method}
Our survey revealed that the explanation method heavily depends on the underlying recommendation algorithm. Moreover, the content-based explanation method received the highest focus, which can be due to the fact that it gives more natural and intuitive explanations to the users based on the features/attributes. An important research question that needs further investigation is how to relate explanatory visualization design with the explanation method. We observed in our survey that node-link diagrams are favorable idioms to explain social recommender systems and tag clouds are mostly used to explain content-based recommender systems. This implies that the explanation method seems to play a crucial role in the selection of the idiom. Therefore, empirical studies are necessary to answer the question ``what kind of visual explanation is more effective for what kind of explanation method?'' \textbf{(R3)}.

\subsection{Explanation Format}
In the design of visual explanations in recommender systems, there are important design decisions to be made at the level of visual encoding and interaction with the explanation. Our review revealed that there is a link between the idiom (how?) and the underlying task (why?) and data (what?) of the visual explanation \textbf{(D1)}. To encode the explanatory visualizations, the most commonly used idioms in our review are node-link diagrams, bar charts, and tag clouds. As the main task of explainable recommendation is to show how a recommended item relates to a user profile, the node-link diagram is the most effective idiom for network-based datasets to enable efficient path identification (e.g., visualize the connection between the recommended item and the user profile) \textbf{(D2)}. Another common task in explainable recommendation is to provide explanations based on the similarity of items or users. In the case of network-based datasets, the node-link diagram is an adequate idiom to cluster items or users based on their similarity. Tag clouds and Venn diagrams are effective idioms for comparing similarities in set-based datasets. And, (grouped) bar charts are intuitive idioms to compare similarities of item/user attributes \textbf{(D3)}.

The tools in our review leveraged visualizations to expose different parts of the recommender system (i.e., input, algorithm's inner workings, and output) and how these parts are interconnected. In the reviewed literature, visualizations exist in different forms and complexity levels and are often combined in highly interactive visualizations. However, users may not be interested in all the information that the visual explanation can produce nor to actively interact with the provided visualizations. The design of visual explanations depends on the goal, target user, and application domain \cite{spinner2019explainer, mohseni2021multidisciplinary}. Different users, classified into AI novices, data experts, AI experts \cite{mohseni2021multidisciplinary} or model novices, model users, model developers \cite{spinner2019explainer}, demand different levels of information and interaction in a visual explanation and there may be negative effects if an explanation is difficult to understand. End-users are not always looking for information-heavy, highly interactive interfaces that are interesting from a visualization perspective, but are too complex for their needs \cite{ooge2022explaining}. One golden rule of data visualization is to maintain simplicity, which also applies to the case of visual explanation as well \cite{kwon2018retainvis}. Thus, it is important to provide visual explanations as simple as possible, yet with enough details to allow users to build accurate mental models of how the recommender system operates, without overwhelming them \textbf{(D4)}. Moreover, to address various users' needs and goals, different variations of the visual explanation with varying level of details need to be considered \textbf{(D5)}. The question that arises here is: What is the tradeoff between the amount of information and interaction in a visual explanation and the level of e.g., satisfaction and trust users develop when they interact with the recommender system, i.e., "not too little and not too much" \cite{kizilcec2016much}. To address this challenge, studies are needed to investigate the right amount of information and interaction in a visual explanation and how different levels of information and interaction (e.g., low, medium, high) affect users’ perception of the visually explainable recommender system \textbf{(R4)}. 

Furthermore, the right amount of information and interaction in a visual explanation should be adapted for different groups of end-users based on their personal characteristics \textbf{(D6)}. Only a few studies addressed the gap between the research about personal characteristics and the research about explanations in recommender systems and provided evidence that personal characteristics have a significant influence on the perception of and interaction with explainable recommender systems \cite{Millecamp2019, Kouki2019}. These studies suggest that categorizing users based on their computational background is not enough since this is not the only influencing criterion on users’ perception of explanations. Thus, there is a need for more profound research on the influencing criteria (e.g., visualization literacy, visual working memory, domain expertise, prior knowledge) and their effects on users’ perception of visually explainable recommender systems as a foundation for the design of effective visual explanations that take personal characteristics into account \textbf{(R5)}. 

Interaction with the explanatory visualizations is an integral part of all surveyed tools. Basic interaction types, such as selecting and filtering data, as well as juxtaposing and coordinating multiple views are the most used ways of interacting with the visual explanations. While the provided interaction types may already provide interesting insights into how the recommender systems operate, combining them with other interaction types would make them more powerful as users can then focus on particular insights or view the explanation information from different angles. The interaction types could be further enhanced, varied, and combined to better support personalized interactive explanation \textbf{(R6)}. To meet different users’ needs, it is crucial to follow a human-in-the-loop approach where the visual explanation acts as a  'conversation partner' and users can interact with, control, and personalize the visual explanation process to build an accurate mental model of the recommender system \textbf{(D7)}. For example, zooming a visualization, collapsing/expanding visual components, tooltips that pop up when hovering or clicking a visualization, and different multiple linked views (MLVs) design patterns, such as small multiples, overview+detail, and multiple views using brushing and linking \cite{Munzner2014} can be leveraged to help users steer the explanation process, personalize the explanation based on their preferences (e.g., whether or not to see the explanation, see different levels of explanation detail, change the explanation viewpoint by focusing on the input, process, and/or output), and change the explanation to answer intelligibility types of interest (e.g., \textit{What}, \textit{Why}, \textit{Why not}, \textit{How}, \textit{What if}, and \textit{How to} questions). Thereby, the visual explanation should support an iterative and layered explanation process driven by the user's needs and goals. For example, the visual explanation could start by providing simple and abstract answers to 'What does the system know about me?' and 'Why was this item recommended to me?' questions. If users are interested in knowing more about the system, they can further interact with the visual explanation to get more detailed answers to the 'How does the system work?' question. Users would also ask follow-up counterfactual 'What if I change my inputs to...?' and 'How to change my inputs to get a different recommendation?' questions to scrutinize their models, explore the consequences of an alternative scenario, and keep closing the gap of understanding. We believe that providing layered visual explanations that follow a "Basic Explanation – Show the Important – Details on Demand" approach can help users iteratively build better mental models of how the system works, without overwhelming them, which was found in recent works to be beneficial for different users \cite{guesmi2021demand, guesmi2023interactive} \textbf{(D8)}.  

We further noted that the current design of the explanatory visualizations in terms of visual encoding and interaction with the explanation is often without reference to well-established design practices from the information visualization field. For instance, Munzner's what-why-how visualization framework (see Section \ref{subsec:vis}), \citet{Ware2012}, and  \citet{Meyer2018} provide practical guidelines on how to get started on developing data visualizations, aligned with what one knows about cognition and perception. The information visualization literature suggests that the \textit{idioms (How?)} depend heavily on the underlying \textit{tasks (Why?)} and \textit{data (What?)} of the visualization and provides guidelines concerning ``which visualization is good for which task'' (Figure~\ref{subfig:infovis-mapping-a}) and ``which visualization is good for which data type'' (Figure~\ref{subfig:infovis-mapping-b}). Designing explainable recommendation with knowledge of information visualization guidelines in mind enables to \textit{get the explanatory visualization right} \textbf{(D9)}. The information visualization literature also provides guidance to design effective visualizations based on a systematic and iterative process, following five major steps: (1) Stakeholders - Who will view the visualization?; (2) Goals / Questions - What are the stakeholders' high-level goals and questions?; (3) Tasks - What are the low-level questions and tasks that will support the high-level goals and questions?; (4) Data - What data is available/required to perform the tasks and what is the type of the data?; and (5) Visualization - What representation of this data will fulfill the tasks for the stakeholders? Following a theory-informed framework for the systematic design of visually explainable recommendation enables to \textit{get the right explanatory visualization} that truly meets user needs \textbf{(D10)}. Future research on explanatory visualizations in recommender systems thus needs to adopt practices, guidelines, and frameworks from the information visualization field to design effective visual explanations \textbf{(R7)}. 

\subsection{Suggestions for Visual Explanation Design}
To summarize the insights gathered through our literature review, we have compiled a set of suggestions for the effective design of visual explanations in recommender systems and future research in this area. 
\begin{enumerate}[label={\bfseries D\arabic*}]
	\item The idiom (how?) depends heavily on the underlying task (why?) and data (what?) of the visual explanation. 
	\item The node-link diagram is an effective idiom to understand the relationship between the input (i.e., user profile) and the output (i.e., recommended item) in a recommender system.	
	\item The node-link diagram / tag cloud / bar chart are simple but effective idioms to provide explanations based on the similarity of items or users in network-based datasets / set-based datasets / item or user attributes.
	\item Provide visual explanations with enough details to help users build accurate mental models of the recommender system, without overwhelming them. 
	\item Consider different variations of a visual explanation with varying level of details to address various users' needs and goals.
	\item Adapt the amount of information and interaction in a visual explanation to match users’ personal characteristics.
	\item Empower users to control and personalize the visual explanation in order to iteratively develop better mental models of the recommender system.
	\item Provide layered visual explanations that follow a "Basic Explanation – Show the Important – Details on Demand" approach.
	\item Design visually explainable recommendation with knowledge of information visualization guidelines in mind to get the explanatory visualization right.
	\item Follow a theory-informed framework for the systematic design of visually explainable recommendation to get the right explanatory visualization.
\end{enumerate}

Further research has to be invested to address the following important questions:
\begin{enumerate}[label={\bfseries R\arabic*}]
	\item What kind of visual explanation is more effective for what kind of explanation aim?
	\item How can visualization be used to open, scrutinize, and explain user models in order to support an effective and efficient personalization and correctability of the recommender system? What are the effects on the perception of and interaction with the recommender system?
	\item What kind of visual explanation is more effective for what kind of explanation method?
	\item What is the right amount of information and interaction in a visual explanation which may foster a better
	understanding, satisfaction, acceptance, and trust in visually explainable recommender systems?
	\item What are the effects of personal characteristics (e.g., visualization literacy, visual working memory, domain expertise, prior knowledge) on the perception of and interaction with visually explainable recommender systems?
	\item How to leverage and combine different interaction types to better support personalized, interactive visual explanation in recommender systems?
	\item How to harness practices, guidelines, and frameworks from the information visualization field to guide the effective design of visual explanations in recommender systems? 
\end{enumerate}

\section{Conclusion} \label{sec:conclusion}
In this paper, we provided an extensive review of the most notable works in the past two decades on explanatory visualizations in recommender systems. We proposed a classification scheme for organizing and clustering existing publications, based on four dimensions: (1) explanation aim, (2) explanation scope, (3) explanation method, and (4) explanation format. We mainly focused on visualization as a display style of explanation. As a result, we derived a set of guidelines which we think might be suggestive for enhancing the design of visual explanation in recommender systems. Moreover, we identified several challenges that remain unaddressed so far, for example, related to research on theory-informed explanatory visualizations in recommender systems and their evaluation.  We hope this survey can provide researchers and practitioners with a comprehensive understanding towards the key aspects of visualization for recommendation explainability as well as better insights on how to effectively design visual explanation in recommender systems. 
\bibliographystyle{ACM-Reference-Format}
\bibliography{references}

\end{document}